\documentclass[12pt]{article}

\usepackage{amsfonts}
\usepackage{amsmath}
\usepackage{epsfig}
\usepackage{latexsym}
\usepackage{graphicx}
\usepackage{epstopdf}
\usepackage{color}
\usepackage{amssymb}
\usepackage{framed}
\numberwithin{equation}{section}
\allowdisplaybreaks

\def\spa#1{\phantom{\fbox{\rule[-#1cm]{0cm}{0cm}}}}

\def\be{\begin{equation}}
\def\ee{\end{equation}}
\def\bea{\begin{eqnarray}}
\def\eea{\end{eqnarray}}
\def\bequ{\begin{equation}}
\def\eequ{\end{equation}}

\def\del{\partial}

\renewcommand{\thefootnote}{\fnsymbol{footnote}}

\newcommand{\eq} {equation}
\newcommand{\eqa} {eqnarray}
\newcommand{\NN} {\nonumber}

\begin{document}
\hfuzz=100pt
\title{Exact relations between M2-brane theories with and without Orientifolds}
\author{
Masazumi Honda\footnote{masazumi.hondaATweizmann.ac.il} 
  \spa{0.5} \\
\\
{\small{\it Department of Particle Physics and Astrophysics,}}
\\ {\small{\it Weizmann Institute of Science, Rehovot 7610001, Israel}} \\
}
\date{\small{December 2015}}

\maketitle
\thispagestyle{empty}
\centerline{}

\begin{abstract}
We study partition functions of low-energy effective theories of M2-branes,
whose type IIB brane constructions include orientifolds.
We mainly focus on circular quiver superconformal Chern-Simons theory on $S^3$,
whose gauge group is $O(2N+1)\times USp(2N)\times \cdots \times O(2N+1)\times USp(2N)$.
This theory is the natural generalization of 
the $\mathcal{N}=5$ ABJM theory with the gauge group $O(2N+1)_{2k} \times USp(2N)_{-k}$.
We find that
the partition function of this type of theory has a simple relation 
to the one of the M2-brane theory without the orientifolds,
whose gauge group is $U(N)\times \cdots \times U(N)$.
By using this relation,
we determine an exact form of the grand partition function of 
the $O(2N+1)_{2} \times USp(2N)_{-1}$ ABJM theory,
where its supersymmetry is expected to be enhanced to $\mathcal{N}=6$.
As another interesting application,
we discuss that
our result gives a natural physical interpretation of 
a relation between the grand partition functions of the $U(N+1)_4 \times U(N)_{-4}$ ABJ theory and $U(N)_2 \times U(N)_{-2}$ ABJM theory,
recently conjectured by Grassi-Hatsuda-Mari\~no.
We also argue that
partition functions of $\hat{A}_3$ quiver theories 
have representations in terms of
an ideal Fermi gas systems associated with $\hat{D}$-type quiver theories and
this leads an interesting relation between certain $U(N)$ and $USp(2N)$ supersymmetric gauge theories.
\end{abstract}
\vfill
\noindent WIS/10/15-NOV-DPPA 

\renewcommand{\thefootnote}{\arabic{footnote}}
\setcounter{footnote}{0}

\newpage
\setcounter{page}{1}
\tableofcontents

\section{Introduction}
In a couple of years,
there is remarkable progress in understanding non-perturbative effects in M-theory through gauge/gravity duality.
Most important tools in this progress are
the supersymmetry localization \cite{Pestun:2007rz,Kapustin:2009kz} and Fermi gas approach \cite{Marino:2011eh}.
These are applied to partition functions in a class of low-energy effective theories of $N$ M2-branes on $S^3$ and
it has turned out that the partition functions are described by an ideal Fermi gas system:
\begin{\eq}
Z(N) =\sum_{\sigma\in S_N} (-1)^\sigma \int d^N x \prod_{j=1}^N \rho (x_j ,x_{\sigma (j)}) ,
\label{eq:Fermi}
\end{\eq}
where 
$\rho$ plays an role of the density matrix in the Fermi gas system.
Thanks to these techniques,
now we know detailed structures of the non-perturbative effects in M-theory on $AdS_4\times S^7/\mathbb{Z}_k$ \cite{Hatsuda:2013oxa,Matsumoto:2013nya,Honda:2014npa,Hatsuda:2013yua},
which is dual to the 3d $\mathcal{N}=6$ superconformal Chern-Simons (CS) theory
known as the ABJ(M) theory \cite{Aharony:2008ug,Aharony:2008gk} via AdS/CFT correspondence 
(see also important earlier works \cite{Marino:2009jd,Drukker:2009hy,Herzog:2010hf,Fuji:2011km,Okuyama:2011su,Hanada:2012si,Hatsuda:2012hm,Hatsuda:2012dt,Awata:2012jb,Hatsuda:2013gj,Honda:2013pea}). 

On the other hand,
we still do not have detailed understanding of the non-perturbative effects 
``beyond ABJ(M) theory'',
namely more general M2-brane theories with less supersymmetry\footnote{
Only exceptions so far are the orbifold ABJM theory and $(2,2)$ model
analyzed in \cite{Honda:2014ica} and \cite{Moriyama:2014gxa,Moriyama:2014nca,Hatsuda:2015lpa}, respectively.
The grand potential for the orbifold ABJM theory
has a simple relation to the one of the ABJM \cite{Honda:2014ica}
and the $(2,2)$ model is expected to be described by topological string on local $D_5$ del Pezzo \cite{Moriyama:2014nca,Hatsuda:2015lpa}.
} (SUSY).
For instance,
it is unclear
whether many attractive features found in the ABJ(M) theory such as the Airy functional behavior \cite{Fuji:2011km,Marino:2011eh}, 
pole cancellation \cite{Hatsuda:2012dt,Hatsuda:2013oxa} and 
correspondence to topological string \cite{Marino:2009jd,Hatsuda:2012hm,Hatsuda:2013oxa}
are universal for general M2-brane theories or accidental for the ABJ(M) theory. 
While the Airy functional behavior has been found 
for a broad class of M2-brane theories \cite{Marino:2011eh,Marino:2012az,Mezei:2013gqa,Assel:2015hsa,Moriyama:2015jsa}
and seems universal \cite{Bhattacharyya:2012ye} (see also \cite{Dabholkar:2014wpa}),
the other features have been found in few examples.
This problem has been addressed in special cases of Imamura-Kimura type theory \cite{Imamura:2008nn},
whose type IIB brane construction
consists of NS5-branes and $(1,k)$-5 branes connected by $N$ D3-branes.
Especially 
the orbifold ABJM theory and $(p,q)$ model \cite{Gaiotto:2008sd} have been studied well in \cite{Mezei:2013gqa,Grassi:2014vwa,Hatsuda:2014vsa,Moriyama:2014gxa,Moriyama:2014nca,Hatsuda:2015lpa}.
Also $\hat{D}$-type quiver theories \cite{Assel:2015hsa,Moriyama:2015jsa}
and $O$ or $USp$ gauge theories with single node \cite{Mezei:2013gqa,Okuyama:2015auc} have been studied (see also \cite{Gulotta:2012yd}).
In order to understand the non-perturbative effects in more detail,
it is very important to investigate 
the non-perturbative effects in  various theories of M2-branes.

In this paper
we consider a generalization along a different direction.
We study partition functions of low-energy effective theories of 
M2-branes on $S^3$,
whose type IIB brane constructions include orientifolds.
We mainly focus on 3d superconformal CS theory of circular quiver type
with the gauge group\footnote{
Recently there appeared a paper \cite{Moriyama:2015asx} 
on arXiv considering a similar physical setup.
This reference mainly considers CS theories of $O(2N)\times USp(2N)$ type,
which differs from our setup of $O(2N+1)\times USp(2N)$ type.
But we also give some comments on the $O(2N)\times USp(2N)$ type in sec.~\ref{sec:O2NUSp}.
} $O(2N+1)\times USp(2N)\times \cdots \times O(2N+1)\times USp(2N)$.
This theory is a natural generalization of 
the $O(2N+1)_{2k} \times USp(2N)_{-k}$ ABJM theory 
with $\mathcal{N}=5$ SUSY \cite{Hosomichi:2008jb,Aharony:2008gk}.
We show that
the $S^3$ partition function of this type of theory is also described 
by an ideal Fermi gas system as in \eqref{eq:Fermi}
and its density matrix $\rho_{O(2N+1)\times USp(2N)}$ takes the following form
\begin{\eq}
\rho_{O(2N+1)\times USp(2N)} (x,y)
=\rho_{U(N)}^{(-)} (x,y) ,
\label{eq:rho_project}
\end{\eq}
where
\begin{\eq}
\rho^{(\pm )} (x,y) = \frac{\rho ( x,y) \pm\rho (x,-y)}{2} .
\end{\eq}
Here $\rho_{U(N)}$ is the density matrix associated 
with the M2-brane theories without the orientifolds,
which are obtained by the replacement $O(2N+1),USp(2N) \rightarrow U(N)$ 
in the orientifold theories.
This indicates that
the density matrix for the orientifold theory is
the projection of the one without the orientifolds.
Introducing the grand canonical partition function by
\begin{\eq}
\Xi [\mu ] =  \sum_N Z(N) e^{\mu N} = {\rm Det}(1+e^\mu \rho ) ,
\end{\eq}
the relation \eqref{eq:rho_project} indicates that
the grand partition function of the orientifold theory
is related to the one of the non-orientifold theory by
\begin{framed}
\begin{\eq}
\Xi_{O(2N+1)\times USp(2N)} [\mu ] =\Xi_{U(N)}^{(-)} [\mu ] ,
\label{eq:main}
\end{\eq}
\end{framed}
\hspace{-2em}
where $\Xi^{(\pm )} [\mu ]$ denotes 
the grand canonical partition function defined by 
$\rho^{(\pm )}$.
This relation
implies that
we can obtain non-perturbative information on the orientifold theory
from the non-orientifold theory.

Here we present two interesting applications of our main result \eqref{eq:main}.
One of them is to determine an exact form of the grand partition function of 
the $O(2N+1)_{2k} \times USp(2N)_{-k}$ ABJM theory with $k=1$,
whose SUSY is expected to be enhanced to $\mathcal{N}=6$ from $\mathcal{N}=5$.
This is achieved by combining our result \eqref{eq:main} with recent results of \cite{Grassi:2014uua,Codesido:2014oua}
and we obtain
\begin{\eqa}
\Xi_{O(2N+1)_2 \times USp(2N)_{-1}} (\mu )
= \Xi_{U(N)_1 \times U(N)_{-1}} ( \mu /2 +\pi i/2 )\cdot  \Xi_{U(N)_1 \times U(N)_{-1}} ( \mu /2 -\pi i/2 ) .
\label{eq:exactk1}
\end{\eqa}
Here $\Xi_{U(N)_1 \times U(N)_{-1}}$ is the grand partition function of the $U(N)_1 \times U(N)_{-1}$ ABJM theory,
whose exact form is conjectured as \cite{Codesido:2014oua}
\begin{\eqa}
&&\Xi_{U(N)_1 \times U(N)_{-1}} (\mu ) \NN\\
&&= \exp{\Biggl[ \frac{3\mu}{4} -\frac{3}{4}\log{2} 
+\mathcal{F}_1 +F_1^{\rm NS} 
-\frac{1}{4\pi^2}\left( \mathcal{F}_0 (\lambda ) 
-\lambda \del_\lambda \mathcal{F}_0 (\lambda ) 
 +\frac{\lambda^2}{2}\del_\lambda^2 \mathcal{F}_0 (\lambda ) \right)
  \Biggr]} \NN\\
&&\times \left(  \vartheta_2 (\bar{\xi}/4 ,\bar{\tau}/4 ) 
+i\vartheta_1 (\bar{\xi}/4 ,\bar{\tau}/4 ) \right) ,
\label{eq:k1ABJM}
\end{\eqa}
where several definitions will be given in sec.~\ref{sec:exact}.

The other application of \eqref{eq:main} is 
to give a natural physical interpretation
of a mysterious relation recently conjectured by Grassi-Hatsuda-Mari\~no \cite{Grassi:2014uua}.
They conjectured 
a relation between the grand partition functions of 
the $U(N+1)_4 \times U(N)_{-4}$ ABJ theory and 
$U(N)_2 \times U(N)_{-2}$ ABJM theory as
\begin{\eq}
\Xi_{U(N)_4 \times U(N+1)_{-4}\rm ABJ} [\mu ] =\Xi_{U(N)_2 \times U(N)_{-2}\rm ABJM}^{(-)} [\mu ] .
\label{eq:GHM}
\end{\eq}
This should be compared with our result \eqref{eq:main} 
for the $\mathcal{N}=5$ ABJM theory:
\begin{\eq}
\Xi_{O(2N+1)_{2k}\times USp(2N)_{-k}} [\mu ] 
=\Xi_{U(N)_{2k}\times U(N)_{-2k}}^{(-)} [\mu ] .
\label{eq:grandABJM}
\end{\eq}
Combining \eqref{eq:GHM} with \eqref{eq:grandABJM},
we find
\begin{\eq}
\Xi_{U(N)_4 \times U(N+1)_{-4}\rm ABJ} [\mu ] 
=\Xi_{O(2N+1)_{2}\times USp(2N)_{-1}} [\mu ] .
\label{eq:enhance}
\end{\eq}
Remarkably this relation is indeed equivalent to the conjecture in \cite{Aharony:2008gk}.
The $\mathcal{N}=5$ ABJM theory is expected to be 
low energy effective theories of $N$ M2-branes probing $\mathbb{C}^4 /\hat{D}_k$ 
with the binary dihedral group $\hat{D}_k$
defined in \eqref{eq:dihedral}.
Since $\mathbb{C}^4 /\hat{D}_k$ for $k=1$ is $\mathbb{C}^4 /\mathbb{Z}_4$, 
moduli of the $O(2N+1)_{2} \times USp(2N)_{-1}$ ABJM theory become
the same as the one of the $U(N+M)_k \times U(N)_{-k}$ ABJ(M) with $k=4$.
Therefore the work \cite{Aharony:2008gk} conjectured that
the $O(2N+1)_{2} \times USp(2N)_{-1}$ ABJM theory 
has the enhanced $\mathcal{N}=6$ SUSY
and equivalent to the $U(N+1)_4 \times U(N)_{-4}$ ABJ theory\footnote{
In order to fix the value of $M$,
we should compare not only the moduli but also discrete torsion \cite{Aharony:2008gk}.
}:
\begin{\eq}
O(2N+1)_{2} \times USp(2N)_{-1}\ \leftrightarrow \ U(N+1)_4 \times U(N)_{-4} ,
\label{eq:isomorphism}
\end{\eq}
which gives\footnote{
This statement has been partially checked by using superconformal index \cite{Cheon:2012be}.
} \eqref{eq:enhance}.
If we assume this, 
then our result \eqref{eq:grandABJM} leads us to the Grassi-Hatsuda-Mari\~no relation \eqref{eq:GHM},
while if we assume \eqref{eq:GHM}, 
then our result \eqref{eq:grandABJM} indicates 
the conjectural equivalence \eqref{eq:enhance}.

We also discuss that
partition functions of $\hat{A}_3$ quiver theories 
have representations in terms of
ideal Fermi gas systems associated with $\hat{D}$-type quivers\footnote{
The papers \cite{Assel:2015hsa,Moriyama:2015jsa}
have written partition functions of $\hat{D}$-type quiver theories
in terms of ideal Fermi gas systems.
it is unclear to us whether their derivation includes $\hat{D}_3$ case.
However their derivation apparently seems to consider $\hat{D}_{n\geq 4}$
and it is unclear to us whether their derivation includes 
the $\hat{D}_3$ case or not.
Hence we explicitly prove this for the $\hat{D}_3$ case.
Even if \cite{Assel:2015hsa,Moriyama:2015jsa} 
did not prove it for the $\hat{D}_3$ case,
our derivation is not essentially new.
} and
this leads an interesting relation 
between certain $U(N)$ and $USp(2N)$ SUSY gauge theories with single node.
The $U(N)$ gauge theory under consideration is
$\mathcal{N}=4$ vector multiplet with one adjoint hyper multiplet and $N_f$ fundamental hyper multiplets,
while
the $USp(2N)$ gauge theory is
$\mathcal{N}=4$ vector multiplet 
with one anti-symmetric hyper multiplet and $N_f$-fundamental hyper multiples.
Regarding these theories,
the work \cite{Okuyama:2015auc} has proposed the equivalence
\begin{\eq}
Z_{U(N)+adj.} (N,N_f =4 ) =Z_{USp(2N)+A}(N,N_f =3 ) .
\label{UequalUSp}
\end{\eq}
This relation is expected from 3d mirror symmetry\footnote{
We thank Kazumi Okuyama for useful discussions on this point.
} \cite{Intriligator:1996ex,deBoer:1996mp}.
It is known that 
the $U(N)$ and $USp(2N)$ theories are equivalent to 
$\hat{A}_{N_f -1}$ and $\hat{D}_{N_f }$ quiver theories without CS terms,
where only one of the vector multiplets 
is coupled to one fundamental hyper multiplet.
Since $\hat{A}_3 =\hat{D}_3$, 
\eqref{UequalUSp} should hold via the 3d mirror symmetries.
In appendix we explicitly prove this relation 
by using the technique in \cite{Assel:2015hsa,Moriyama:2015jsa}.

This paper is organized as follows.
In sec.~\ref{sec:ABJM},
we consider the $\mathcal{N}=5$ ABJM theory 
with the gauge group $O(2N+1)_{2k}\times USp(2N)_{-k}$.
In sec.~\ref{sec:general},
we generalize our analysis in sec.~\ref{sec:ABJM}
to more general quiver gauge theories.
We also identify quantum mechanical operators in ideal Fermi gas systems
naturally corresponding to orientifolds in type IIB brane constructions.
As interesting examples,
we deal with orientifold projections of 
the $(p,q)$ model and orbifold ABJM theory.
Section \ref{sec:conclusion} is devoted to conclusion and discussions.
In appendix,
we explicitly prove the equivalence \eqref{UequalUSp}.

\newpage
\section{$O(2N+1)_{2k} \times USp(2N)_{-k}$ ABJM theory}
\label{sec:ABJM}
In this section we consider the $\mathcal{N}=5$ ABJM theory 
with the gauge group $O(2N+1)_{2k}\times USp(2N)_{-k}$.
We will generalize our analysis in this section
to more general theory in next section.

\subsection{Orientifold ABJM theory as a Fermi gas}
\label{sec:rhoABJM}
\begin{table}[t]
\begin{center}
  \begin{tabular}{|c|c |  }
  \hline    Multiplet                                            & One-loop determinant    \\\hline
 $\mathcal{N}=2$  $O(2N+1)$ vector multiplet  &     
$ \prod_{i<j} \Bigl[ 2\sinh{\frac{\mu_i -\mu_j}{2}}\cdot 2\sinh{\frac{\mu_i + \mu_j}{2}} \Bigr]^2 \prod_{j=1}^N 4\sinh^2{\frac{\mu_j}{2}}$   \\\hline
 $\mathcal{N}=2$  $USp(2N)$ vector multiplet  &     
$\prod_{i<j} \Bigl[  2\sinh{\frac{\nu_i -\nu_j}{2}}\cdot 2\sinh{\frac{\nu_i + \nu_j}{2}} \Bigr]^2 \prod_{j=1}^N 4\sinh^2{\nu_j}$
   \\\hline
$O(2N+1)\times USp(2N)$ bi-fund. chiral mult. &
$\left( \prod_{i,j}\Bigl[  2\cosh{\frac{\mu_i -\nu_j}{2}}\cdot 2\cosh{\frac{\mu_i + \nu_j}{2}}
\Bigr]  \prod_j 2\cosh{\frac{\nu_j}{2}} \right)^{-1}$ \\\hline
  \end{tabular}
\end{center}
\caption{One-loop determinant of each multiplet in the localization of the $O(2N+1)_{2k} \times USp(2N)_{-k}$ ABJM theory on $S^3$. 
}
\label{tab:1loop}
\end{table}
Thanks to the localization \cite{Kapustin:2009kz},
the partition function of 
the $O(2N+1)_{2k} \times USp(2N)_{-k}$ ABJM theory on $S^3$ 
can be written as\footnote{
Note that the $O(2N+1)_{2k} \times USp(2N)_{-k}$ $\mathcal{N}=5$ ABJM theory
has only one bi-fundamental hyper multiplet
since one of two bi-fundamental hyper multiplets 
in the $\mathcal{N}=6$ ABJM theory
is projected out by the orientifold projection \cite{Hosomichi:2008jb,Aharony:2008gk}.
In the localization formula, 
the $O\times USp$ bi-fundamental chiral multiplet (with R-charge 1/2)
behaves like ``half'' of the hyper multiplet because of the group structure.
} (see tab.~\ref{tab:1loop} for detail)
\begin{eqnarray}
Z_{\mathcal{N}=5 \rm ABJM}(N) \nonumber 
&=& \frac{1}{2^{2N}N!^2} \int \frac{d^N \mu}{(2\pi )^N}
 \frac{d^N \nu}{(2\pi )^N} 
e^{\frac{ik}{2\pi}\sum_{j=1}^N (\mu_j^2 -\nu_j^2 )}  
\prod_{j=1}^N 4\sinh^2{\frac{\mu_j}{2}}\cdot 4\sinh^2{\nu_j} \NN\\
&&\times
\frac{\prod_{i<j} \Bigl[ 
2\sinh{\frac{\mu_i -\mu_j}{2}}\cdot 2\sinh{\frac{\mu_i + \mu_j}{2}}\cdot
2\sinh{\frac{\nu_i -\nu_j}{2}}\cdot 2\sinh{\frac{\nu_i + \nu_j}{2}}
\Bigr]^2}
{\prod_{i,j}\Bigl[ 
2\cosh{\frac{\mu_i -\nu_j}{2}}\cdot 2\cosh{\frac{\mu_i + \nu_j}{2}} \Bigr]^2 
\prod_j 4\cosh^2{\frac{\nu_j}{2}}} .
\end{eqnarray}
Now we write
$Z_{\mathcal{N}=5 \rm ABJM}$ in terms of an ideal Fermi gas 
as in circular quiver $U(N)$ SUSY gauge theories \cite{Marino:2011eh,Okuyama:2011su}.
For this purpose we use the Cauchy determinant-like formula \cite{Mezei:2013gqa}
\begin{\eqa}
&&\frac{\prod_{i<j} \Bigl[ 
2\sinh{\frac{\mu_i -\mu_j}{2}}\cdot 2\sinh{\frac{\mu_i + \mu_j}{2}}\cdot
2\sinh{\frac{\nu_i -\nu_j}{2}}\cdot 2\sinh{\frac{\nu_i + \nu_j}{2}}
\Bigr]}
{\prod_{i,j}\Bigl[ 
2\cosh{\frac{\mu_i -\nu_j}{2}}\cdot 2\cosh{\frac{\mu_i + \nu_j}{2}} \Bigr]} \NN\\
&&  = \sum_{\sigma\in S_N} (-1)^\sigma \prod_j 
     \frac{1}{ 2\cosh{\left( \frac{\mu_j -\nu_{\sigma (j)}}{2} \right) } \cdot
                 2\cosh{\left( \frac{\mu_j +\nu_{\sigma (j)}}{2} \right) }  } ,
\label{eq:Cauchy}
\end{\eqa}
and rewrite the partition function as
\begin{eqnarray}
&&Z_{\mathcal{N}=5 \rm ABJM}(N) \\\nonumber 
&&= \frac{1}{2^{2N}N!} \sum_{\sigma\in S_N} (-1)^\sigma
 \int \frac{d^N \mu}{(2\pi )^N} \frac{d^N \nu}{(2\pi )^N} 
e^{\frac{ik}{2\pi}\sum_{j=1}^N (\mu_j^2 -\nu_j^2 )}  \prod_{j=1}^N 4\sinh^2{\frac{\mu_j}{2}}\cdot 4\sinh^2{\nu_j} \NN\\
&&\times
\prod_j   \frac{1}{2\cosh{\left( \frac{\mu_j -\nu_j}{2} \right) } \cdot
                 2\cosh{\left( \frac{\mu_j +\nu_j}{2} \right) } \cdot
                 2\cosh{\left( \frac{\mu_j -\nu_{\sigma (j)}}{2} \right) } \cdot
                 2\cosh{\left( \frac{\mu_j +\nu_{\sigma (j)}}{2} \right) }  4\cosh^2{\frac{\nu_j}{2}} } \NN\\ 
&=& \frac{1}{N!} \sum_{\sigma\in S_N} (-1)^\sigma \int  d^N \mu   \prod_j \rho_{\mathcal{N}=5 \rm ABJM} (\mu_j ,\mu_{\sigma (j)} ) ,
\end{eqnarray}
where
\begin{\eqa}
\rho_{\mathcal{N}=5 \rm ABJM} (x,y) 
&=& \frac{1}{2\pi k'}\sinh{\frac{x}{2k'}} \sinh{\frac{y}{2k'}}   \NN\\
&&\int \frac{d\nu}{2\pi k'}    \frac{\sinh^2{\frac{\nu}{k'}}\cdot e^{\frac{i}{4\pi k'}(x^2 -\nu^2 )}}
 {2\cosh{\left( \frac{x -\nu}{2k'} \right) } \cdot 2\cosh{\left( \frac{x +\nu}{2k'} \right) } \cdot
                 2\cosh{\left( \frac{\nu -y}{2k'} \right) } \cdot
                  2\cosh{\left( \frac{\nu +y}{2k'} \right) } \cdot
                  \cosh^2{\frac{\nu}{2k'}}  } ,    \NN\\
\end{\eqa}
with $k'=2k$.
This equation tells us that
the partition function of the $\mathcal{N}=5$ ABJM theory is described by the ideal Fermi gas system 
with the density matrix $\rho_{\mathcal{N}=5 \rm ABJM} (x,y)$.
We regard $\rho_{\mathcal{N}=5 \rm ABJM} (x,y)$ 
as the matrix element of a quantum mechanical operator 
as in \cite{Marino:2011eh}, 
\begin{\eq}
\rho_{\mathcal{N}=5 \rm ABJM} (x,y) 
= \frac{1}{\hbar} \langle x| \hat{\rho}_{\mathcal{N}=5 \rm ABJM}(\hat{Q} ,\hat{P}) |y\rangle ,
\end{\eq}
where
\begin{\eq}
[\hat{Q} ,\hat{P} ] = i\hbar ,\quad \hbar =2\pi k' =4\pi k .
\label{eq:hbar}
\end{\eq}
The operator $\hat{\rho}_{\mathcal{N}=5 \rm ABJM} $ is defined as
\begin{\eqa}
\hat{\rho}_{\mathcal{N}=5 \rm ABJM} (\hat{Q} ,\hat{P}) 
&=&\frac{1}{4} e^{\frac{i}{2\hbar}\hat{Q}^2} \frac{1-\hat{R}}{2\cosh{\frac{\hat{P}}{2}}} \frac{1}{2\sinh{\frac{\hat{Q}}{2k'}}}
  e^{-\frac{i}{2\hbar}\hat{Q}^2} \frac{\sinh^2{\frac{\hat{Q}}{k}} }{2\sinh{\frac{\hat{Q}}{2k'}}  \cosh^2{\frac{\hat{Q}}{2k'}}}
 \frac{1-\hat{R}}{2\cosh{\frac{\hat{P}}{2}}}  \NN\\
&=&e^{\frac{i}{2\hbar}\hat{Q}^2} \frac{1}{2\cosh{\frac{\hat{P}}{2}}} \frac{1-\hat{R}}{2} 
 e^{-\frac{i}{2\hbar}\hat{Q}^2}  \frac{1}{2\cosh{\frac{\hat{P}}{2}}} \frac{1-\hat{R}}{2}  ,
\label{eq:rhoABJM}
\end{\eqa}
where $\hat{R}|x\rangle = |-x\rangle$ and we have used
\begin{\eqa}
\frac{2\sinh{\frac{x}{2k'}}\cdot 2\sinh{\frac{y}{2k'}}}{2\cosh{\frac{x -y}{2k'}}\cdot 2\cosh{\frac{x +y}{2k'}}}
=\langle x| \frac{1-\hat{R}}{2\cosh{\frac{\hat{P}}{2}}}  |y\rangle .
\label{eq:minus}
\end{\eqa}
By using the operator equations\footnote{
Note also $\hat{R}f(\hat{Q})=f(-\hat{Q})\hat{R}$, $\hat{R}f(\hat{P})=f(-\hat{P})\hat{R}$ and $((1-\hat{R})/2)^2 =(1-\hat{R})/2$.
}
$e^{\frac{i}{2\hbar}\hat{Q}^2} f(\hat{P})e^{-\frac{i}{2\hbar}\hat{Q}^2}=f(\hat{P}-\hat{Q})$
and $e^{\frac{i}{2\hbar}\hat{P}^2} g(\hat{Q})e^{-\frac{i}{2\hbar}\hat{P}^2}=g(\hat{Q}+\hat{P})$,
we simplify $\hat{\rho}_{\mathcal{N}=5 \rm ABJM}$ as
\begin{\eq}
\hat{\rho}_{\mathcal{N}=5 \rm ABJM} (\hat{Q} ,\hat{P}) 
= \frac{1}{2\cosh{\frac{\hat{Q}-\hat{P}}{2}}}  \frac{1}{2\cosh{\frac{\hat{P}}{2}}} \frac{1-\hat{R}}{2}  
=  e^{\frac{i}{2\hbar}\hat{P}^2}  \frac{1}{2\cosh{\frac{\hat{Q}}{2}}} e^{-\frac{i}{2\hbar}\hat{P}^2}  \frac{1}{2\cosh{\frac{\hat{P}}{2}}} \frac{1-\hat{R}}{2}  .
\end{\eq}
Performing the similarity transformation
\begin{\eq}
\hat{\rho}_{\mathcal{N}=5 \rm ABJM} (\hat{Q} ,\hat{P}) 
\rightarrow 
\left(  \sqrt{2\cosh{\frac{\hat{Q}}{2}}} e^{\frac{i}{2\hbar}\hat{P}^2}\right)  
\hat{\rho}_{\mathcal{N}=5 \rm ABJM} (\hat{Q} ,\hat{P}) 
\left(  \sqrt{2\cosh{\frac{\hat{Q}}{2}}} e^{\frac{i}{2\hbar}\hat{P}^2}\right)^{-1} ,
\label{eq:rhoN5ABJ}
\end{\eq}
we obtain the following highly simplified expression
\begin{\eq}
\hat{\rho}_{\mathcal{N}=5 \rm ABJM}(\hat{Q} ,\hat{P}) 
= \frac{1}{\left( 2\cosh{\frac{\hat{Q}}{2}} \right)^{1/2}}   \frac{1}{2\cosh{\frac{\hat{P}}{2}}} 
\frac{1}{\left( 2\cosh{\frac{\hat{Q}}{2}} \right)^{1/2}}  \frac{1-\hat{R}}{2} .
\end{\eq}
Recalling that $\hat{\rho}$ for the $\mathcal{N}=6$ ABJM theory 
with the gauge group $U(N)_{2k} \times U(N)_{-2k}$
is given by\footnote{
Note that the definition of $\hbar$ in \eqref{eq:hbar} is slightly different from the one usually used 
in Fermi gas systems associated with $U(N)$ CS theories.
}  
\begin{\eq}
\hat{\rho}_{\mathcal{N}=6\rm ABJM}(\hat{Q} ,\hat{P}) 
=   \frac{1}{\left( 2\cosh{\frac{\hat{Q}}{2}} \right)^{1/2}}  
 \frac{1}{2\cosh{\frac{\hat{P}}{2}}} 
  \frac{1}{\left( 2\cosh{\frac{\hat{Q}}{2}} \right)^{1/2}}  ,
\end{\eq}
we finally obtain
\begin{\eq}
\hat{\rho}_{\mathcal{N}=5 \rm ABJM}(\hat{Q} ,\hat{P}) 
= \hat{\rho}_{\mathcal{N}=6 \rm ABJM}(\hat{Q} ,\hat{P}) \frac{1-\hat{R}}{2} .
\label{eq:rho_final}
\end{\eq}
This indicates that
the density matrix operator $\hat{\rho}_{\mathcal{N}=5 \rm ABJM}$ 
of the $\mathcal{N}=5$ ABJM theory is
the projection of the one of the $\mathcal{N}=6$ ABJM theory.
Since the $\mathcal{N}=5$ ABJM theory is 
the orientifold projection of the $\mathcal{N}=6$ ABJM theory,
presumably the operation of $(1-\hat{R})/2$ 
to $\hat{\rho}_{\mathcal{N}=6 \rm ABJM}$ corresponds 
to the orientifold projection.
It is interesting 
if one can understand this relation more precisely.

\subsubsection*{Remarks}
\begin{enumerate}
\item
The representation \eqref{eq:rho_final} of $\hat{\rho}_{\mathcal{N}=5 \rm ABJM}$ gives the matrix element
\begin{\eq}
\rho_{\mathcal{N}=5 \rm ABJM} (x,y)
=\frac{1}{2}
\frac{1}{\sqrt{2\cosh{\frac{x}{2}}}}
\frac{\sinh{\frac{x}{4k}}\sinh{\frac{y}{4k}}}{\cosh{\frac{x -y}{4k}}\cosh{\frac{x +y}{4k}}} 
\frac{1}{\sqrt{2\cosh{\frac{y}{2}}}} .
\end{\eq}
This gives the following representation of the partition function
\begin{\eq}
Z_{\mathcal{N}=5 \rm ABJM} (N,k)
= \frac{1}{2^{N}N!} \int \frac{d^N x}{(2\pi )^N} 
 \prod_{j=1}^N \frac{4\sinh^2{\frac{x_j}{2k}}}{2\cosh{x_j} } 
\frac{\prod_{i<j} \Bigl[ 
2\sinh{\frac{x_i -x_j}{2k}}\cdot 2\sinh{\frac{x_i + x_j}{2k}}\Bigr]^2}
{\prod_{i,j}\Bigl[  2\cosh{\frac{x_i -x_j}{2k}}\cdot 2\cosh{\frac{x_i + x_j}{2k}} \Bigr] }  ,
\label{eq:N5ABJM2}
\end{\eq}
where we have rescaled as $x\rightarrow 2x$.
Let us compare this with the 
partition function of
the $USp(2N)$ 
gauge theory with $\mathcal{N}=4$ vector multiplet, 
one symmetric hyper multiplet and $N_f$-fundamental hyper multiples\footnote{
When we go to the last line from the second line,
we have used $\sinh^2{\mu_j} = 4\sinh^2{\frac{\mu_j}{2}} \cosh^2{\frac{\mu_j}{2}} $.
} 
(called $USp+S$ theory in \cite{Mezei:2013gqa}):
\begin{\eqa}
&&Z_{USp+S}(N, N_f )  \NN\\
&&= \frac{1}{2^N N!}
\int \frac{d^N \mu}{(2\pi )^N}
  \prod_{j=1}^N \frac{4\sinh^2{\mu_j}}{2\cosh{\mu_j} \left( 2\cosh{\frac{\mu_j}{2}} \right)^{2N_f}}
  \frac{\prod_{i<j} \Bigl[ 2\sinh{\frac{\mu_i -\mu_j}{2}}\cdot 2\sinh{\frac{\mu_i + \mu_j}{2}} \Bigr]^2 }
   {\prod_{i,j} \cosh{\frac{\mu_i -\mu_j}{2}}\cdot \cosh{\frac{\mu_i + \mu_j}{2}} }  \NN\\
&&= \frac{1}{2^N N!}
\int \frac{d^N \mu}{(2\pi )^N}
  \prod_{j=1}^N \frac{4\sinh^2{\frac{\mu_j}{2}}}{2\cosh{\mu_j} \left( 2\cosh{\frac{\mu_j}{2}} \right)^{2N_f -2}}
  \frac{\prod_{i<j} \Bigl[ 2\sinh{\frac{\mu_i -\mu_j}{2}}\cdot 2\sinh{\frac{\mu_i + \mu_j}{2}} \Bigr]^2 }
   {\prod_{i,j} \cosh{\frac{\mu_i -\mu_j}{2}}\cdot \cosh{\frac{\mu_i + \mu_j}{2}} }  .\NN\\
\end{\eqa}
Comparing this with \eqref{eq:N5ABJM2},
we easily see that
the $\mathcal{N}=5$ ABJM theory with $k=1$ agrees 
with\footnote{
We can also compare this with the $O(2N+1)$ gauge theory with $\mathcal{N}=4$ vector multiplet, 
one symmetric hyper multiplet and $N_f$-fundamental hyper multiples ($O(2N+1)+S$ theory).
Because of
$Z_{O(2N+1)+S}(N, N_f  )=Z_{USp+S}(N, N_f -2 )$,
the relation \eqref{eq:USp+S} also shows $Z_{\mathcal{N}=5 \rm ABJM} (N,k=1) = Z_{O(2N+1)+S}(N, N_f =3 )$.
} the $USp+S$ theory with $N_f =1$:
\begin{\eq}
Z_{\mathcal{N}=5 \rm ABJM} (N,k=1) = Z_{USp+S}(N, N_f =1 ) . 
\label{eq:USp+S}
\end{\eq}
It is interesting
if one can understand 
this relation by the brane constructions.
Note that
this result is essentially the same as the recent result 
in \cite{Okuyama:2015auc},
which has shown 
the equivalence between the grand partition function of
the $USp+S$ theory with $N_f =1$ and $\Xi^{(-)}$ part of the $U(N)_2 \times U(N)_{-2}$ ABJM theory.
Because of \eqref{eq:rho_final},
our result is equivalent to this result: 
$\Xi_{O(2N+1)_2\times USp(2N)_{-1}}  =\Xi_{U(N)_2 \times U(N)_{-2}}^{(-)}  =\Xi_{USp+S} (N_f =1)$.

\item
When we identify the quantum mechanical operator \eqref{eq:rhoABJM} 
associated with $\rho_{\mathcal{N}=5 \rm ABJM} (x,y)$,
we could use the following identity once or twice instead of \eqref{eq:minus},
\begin{\eqa}
\frac{2\cosh{\frac{x}{2k}}\cdot 2\cosh{\frac{y}{2k}}}{2\cosh{\frac{x -y}{2k}}\cdot 2\cosh{\frac{x +y}{2k}}}
=\langle x| \frac{1+\hat{R}}{2\cosh{\frac{\hat{P}}{2}}}  |y\rangle .
\label{eq:idplus}
\end{\eqa}
Then the partition function $Z_{\mathcal{N}=5 \rm ABJM}$ is described by different representations of $\hat{\rho}$.
If we use this identity and \eqref{eq:minus} just once by once,
then we find
\begin{\eqa}
\hat{\rho}_{\mathcal{N}=5 \rm ABJM}'
&=&\frac{1}{4} \frac{e^{\frac{i}{2\hbar}\hat{Q}^2}}{2\cosh{\frac{\hat{Q}}{2k}}} \frac{1+\hat{R}}{2\cosh{\frac{\hat{P}}{2}}} 
\frac{  e^{-\frac{i}{2\hbar}\hat{Q}^2}}{2\cosh{\frac{\hat{Q}}{2k}}} 
  \frac{1}{2\sinh{\frac{\hat{Q}}{2k}}}
  \frac{ \sinh^2{\frac{\hat{Q}}{k}}}{ \cosh^2{\frac{\hat{Q}}{2k}}}
\frac{1-\hat{R}}{2\cosh{\frac{\hat{P}}{2}}} \frac{1}{2\sinh{\frac{\hat{Q}}{2k}}} \NN\\
&=&\frac{1}{4} \frac{1}{2\cosh{\frac{\hat{Q}}{2k}}} \frac{1+\hat{R}}{2\cosh{\frac{\hat{P}-\hat{Q}}{2}}} 
   \frac{\sinh{\frac{\hat{Q}}{k}}}{\cosh^2{\frac{\hat{Q}}{2k}}} \frac{1-\hat{R}}{2\cosh{\frac{\hat{P}}{2}}} \frac{1}{2\sinh{\frac{\hat{Q}}{2k}}} ,
\end{\eqa}
while
if we use \eqref{eq:idplus} twice,
then we get
\begin{\eq}
\hat{\rho}_{\mathcal{N}=5 \rm ABJM}''
= \frac{1}{4} \frac{e^{\frac{i}{2\hbar}\hat{Q}^2}}{2\cosh{\frac{\hat{Q}}{2k}}} \frac{1+\hat{R}}{2\cosh{\frac{\hat{P}}{2}}} 
  \frac{e^{-\frac{i}{2\hbar}\hat{Q}^2}}{2\cosh{\frac{\hat{Q}}{2k}}}
\frac{1}{2\cosh{\frac{\hat{Q}}{2k}}}  \frac{ \sinh^2{\frac{\hat{Q}}{k}}}{\cosh^2{\frac{\hat{Q}}{2k}}} 
\frac{1+\hat{R}}{2\cosh{\frac{\hat{P}}{2}}} 
\frac{1}{2\cosh{\frac{\hat{Q}}{2k}}} .
\end{\eq}
To summarize,
we have four different representations of $\hat{\rho}$ 
to describe the same partition function $Z_{\mathcal{N}=5 \rm ABJM}$:
\begin{\eq}
\hat{\rho}_{\mathcal{N}=5 \rm ABJM}
=\frac{1}{4}e^{\frac{i}{2\hbar}\hat{Q}^2} f_\pm (\hat{Q}) \frac{1\pm\hat{R}}{2\cosh{\frac{\hat{P}}{2}}} f_\pm (\hat{Q})
  e^{-\frac{i}{2\hbar}\hat{Q}^2}\cdot \frac{\sinh^2{\frac{\hat{Q}}{k}}}{\cosh^2{\frac{\hat{Q}}{2k}}} f_\pm (\hat{Q})
\frac{1\pm\hat{R}}{2\cosh{\frac{\hat{P}}{2}}} f_\pm (\hat{Q}) ,
\end{\eq}
where we can freely choose ``$+$'' or ``$-$'' 
at every ``$f_\pm (1\pm \hat{R})f_\pm$'' and
$f_\pm$ is given by
\begin{\eq}
f_+ (Q) = \frac{1}{2\cosh{\frac{Q}{2k}}} ,\quad 
f_- (Q) = \frac{1}{2\sinh{\frac{Q}{2k}}} .
\end{\eq}
In this paper we always choose ``$-$''
since taking ``$-$'' seems 
technically simpler. 
\end{enumerate}

\subsection{Exact grand partition function for $k=1$}
\label{sec:exact}
Here we find the exact form of the grand partition function of the $O(2N+1)_{2k} \times USp(2N)_{-k}$ ABJM theory for $k=1$.
Grassi, Hatsuda and Mari\~no conjectured \cite{Grassi:2014uua}
\begin{\eq}
\Xi_{U(N)_2 \times U(N)_{-2}}^{(-)} (\mu )
= \Xi_{U(N)_1 \times U(N)_{-1}} ( \mu /2 +\pi i/2 )\cdot  \Xi_{U(N)_1 \times U(N)_{-1}} ( \mu /2 -\pi i/2 ) .
\end{\eq}
Combining this with our result \eqref{eq:rho_final},
we immediately find \eqref{eq:exactk1}
\begin{\eqa *}
\Xi_{O(2N+1)_2 \times USp(2N)_{-1}} (\mu )
&=& \Xi_{U(N)_2 \times U(N)_{-2}}^{(-)} (\mu ) \NN\\
&=& \Xi_{U(N)_1 \times U(N)_{-1}} ( \mu /2 +\pi i/2 ) \cdot  \Xi_{U(N)_1 \times U(N)_{-1}} ( \mu /2 -\pi i/2 ) .
\end{\eqa *}
The exact form \eqref{eq:k1ABJM} of the grand partition function $\Xi_{U(N)_1 \times U(N)_{-1}}$
was proposed as \cite{Codesido:2014oua}
\begin{\eqa *}
&&\Xi_{U(N)_1 \times U(N)_{-1}} (\mu )\NN\\
&&= \exp{\Biggl[ \frac{3\mu}{4} -\frac{3}{4}\log{2} +\mathcal{F}_1 +F_1^{\rm NS} 
-\frac{1}{4\pi^2}\left( \mathcal{F}_0 (\lambda ) -\lambda \del_\lambda \mathcal{F}_0 (\lambda ) 
 +\frac{\lambda^2}{2}\del_\lambda^2 \mathcal{F}_0 (\lambda ) \right) \Biggr]} \NN\\
&&\ \ \ \times \left(  \vartheta_2 (\bar{\xi}/4 ,\bar{\tau}/4 ) +i\vartheta_1 (\bar{\xi}/4 ,\bar{\tau}/4 ) \right) ,
\end{\eqa *}
where $\vartheta_{1,2}$ is the Jacobi theta function\footnote{
Their definitions are
\[
\vartheta_1 (v,\tau ) 
=\sum_{n\in\mathbb{Z}} (-1)^{n-1/2} e^{\pi i(n+1/2)^2 \tau +2\pi i (n+1/2)v} ,\quad
\vartheta_2 (v,\tau ) 
=\sum_{n\in\mathbb{Z}}  e^{\pi i(n+1/2)^2 \tau +2\pi i (n+1/2)v} .
\]
} and\footnote{See \cite{Codesido:2014oua} for details.}
\begin{\eqa}
&&\lambda 
= \frac{\kappa^2}{8\pi}\ _3 F_2 \left( \frac{1}{2},\frac{1}{2},\frac{1}{2}; 1,\frac{3}{2}; -\frac{\kappa^4}{16} \right) ,\quad
\kappa =e^\mu \NN\\
&&\del_\lambda \mathcal{F}_0 (\lambda )
= \frac{\kappa^2}{4} G_{3,3}^{2,3} \left( \left. \begin{matrix} \frac{1}{2}, & \frac{1}{2}, & \frac{1}{2} \cr 0,&0,& -\frac{1}{2} \end{matrix} \right| -\frac{\kappa^4}{16}\right)
 +\frac{\pi^2 i\kappa^2}{2}\ _3 F_2 \left( \frac{1}{2},\frac{1}{2},\frac{1}{2}; 1,\frac{3}{2}; -\frac{\kappa^4}{16} \right) ,\NN\\
&& \del_\lambda^2 \mathcal{F}_0 (\lambda ) = -8\pi^3 i\bar{\tau} ,\quad
\bar{\xi} =\frac{i}{4\pi^3} \left( \lambda\del_\lambda^2 \mathcal{F}_0 (\lambda ) -\del_\lambda \mathcal{F}_0 (\lambda ) \right) ,\NN\\
&& \mathcal{F}_1 = -\log{\eta (2\bar{\tau})} -\frac{1}{2}\log{2},\quad
F_1^{\rm NS} = \frac{1}{12}\log{e^{-4\mu}} -\frac{1}{24}\log{1+16e^{-4\mu}}.
\end{\eqa}
In terms of \eqref{eq:exactk1},
we can explicitly write the exact form of the grand partition function of 
the $O(2N+1)_{2} \times USp(2N)_{-1}$ ABJM theory.

\subsection{Grassi-Hatsuda-Mari\~no exact functional relation from geometry}
Grassi, Hatsuda and Mari\~no conjectured 
the relation \eqref{eq:GHM} on the grand canonical partition function of the ABJ theory \cite{Grassi:2014uua}:
\[
\Xi_{U(N+1)_4 \times U(N)_{-4} } (\mu ) = \Xi_{U(N)_2 \times U(N)_{-2}}^{(-)} (\mu)  .
\]
Physical interpretation of this relation has been unclear and
therefore this relation has been considered as accidental. 
Now we give a physical interpretation on this relation.
Let us compare this result with our result \eqref{eq:main}:
\[
\Xi_{O(2N+1)_{2k}\times USp(2N)_{-k}} [\mu ] =\Xi_{U(N)_{2k}\times U(N)_{-2k}}^{(-)} [\mu ] .
\]
Plugging \eqref{eq:grandABJM} into \eqref{eq:GHM} leads us to
\[
\Xi_{U(N)_4 \times U(N+1)_{-4}\rm ABJ} [\mu ] 
=\Xi_{O(2N+1)_{2}\times USp(2N)_{-1}} [\mu ] .
\]
This relation is equivalent to the conjecture in \cite{Aharony:2008gk}.
The $O(2N+1)_{2k}\times USp(2N)_{-k}$ ABJM theory 
is expected to be low energy effective theories of $N$ M2-branes probing $\mathbb{C}^4 /\hat{D}_k$ 
with the binary dihedral group $\hat{D}_k$,
whose action to the complex coordinate $(z_1 ,z_2 ,z_3 , z_4)$ 
of $\mathbb{C}^4$ is 
\begin{\eq}
(z_1 ,z_2 ,z_3 ,z_4) 
\sim e^{\frac{\pi i}{k}}(z_1 ,z_2 ,z_3 ,z_4)
\sim (iz_2^\ast , -iz_1^\ast ,iz_4^\ast ,-iz_3^\ast ) .
\label{eq:dihedral}
\end{\eq}
Since $\mathbb{C}^4 /\hat{D}_k$ for $k=1$ is $\mathbb{C}^4 /\mathbb{Z}_4$, 
the moduli of the $O(2N+1)_{2} \times USp(2N)_{-1}$ ABJM theory become
the same as the one of the $U(N+M)_4 \times U(N)_{-4}$ ABJ(M) theory.
Therefore the work \cite{Aharony:2008gk} conjectured that
the $O(2N+1)_{2} \times USp(2N)_{-1}$ ABJM theory has 
$\mathcal{N}=6$ SUSY
and equivalent to the $U(N+1)_4 \times U(N)_{-4}$ ABJ theory 
(see \cite{Cheon:2012be} for the test by superconformal index):
\[
O(2N+1)_{2} \times USp(2N)_{-1}\ \leftrightarrow \ U(N+1)_4 \times U(N)_{-4} .
\]
If we assume this, 
then our result \eqref{eq:main} leads us to the Grassi-Hatsuda-Mari\~no relation \eqref{eq:GHM},
while if we assume the Grassi-Hatsuda-Mari\~no relation \eqref{eq:GHM}, 
then our result \eqref{eq:main} indicates the conjecture \eqref{eq:isomorphism}.

\section{Generalization}
\label{sec:general}
In this section we generalize our analysis in sec.~\ref{sec:ABJM}
to a class of CS theory,
which is circular quiver with the gauge group $[O(2N+1 ) \times USp(2N) ]^r$
and bi-fundamental chiral multiplets one by one between nearest neighboring pairs of the gauge groups.

\subsection{Fermi gas formalism}
\label{sec:generalFermi}
Let us consider the circular quiver CS theory
with the gauge group
$O(2N+1)_{2k_1}\times USp (2N)_{k_1'} \times \cdots \times O(2N+1)_{2k_r}\times USp (2N)_{k_r'}$,
where $O(2N+1)_{2k_a}$ and $USp(2N+1)_{2k_a'}$ are coupled to $N_f^{(a)}$ and $N_f^{\prime (a)}$ fundamental hyper multiplets, respectively. 
We parametrize the CS levels $k_a ,k_a'$ as
$k_a = k n_a , k_a' = k n_a'$
with rational numbers $n_a$ and $n_a'$.
Applying the localization, the partition function becomes \cite{Kapustin:2009kz}
\begin{eqnarray}
&& Z_{O(2N+1)\times USp(2N)}(N) \nonumber \\
&&= \frac{1}{2^{2rN}N!^{2r}} \int \prod_{a=1}^r \frac{d^N \mu^{(a)}}{(2\pi )^N} \frac{d^N \nu^{(a)}}{(2\pi )^N} 
   \prod_{j=1}^N 4\sinh^2{\frac{\mu_j^{(a)}}{2}} f^{(a)}(\mu_j^{(a)} ) \cdot \frac{\sinh^2{\nu_j^{(a)}}}{\cosh^2{\frac{\nu_j^{(a)}}{2}}}  
f^{\prime (a)}(\nu_j^{(a)} )   \NN\\
&&\ \ \ \times
\frac{\prod_{i<j} \Bigl[ 
2\sinh{\frac{\mu_i^{(a)} -\mu_j^{(a)}}{2}}\cdot 2\sinh{\frac{\mu_i^{(a)} + \mu_j^{(a)}}{2}}\cdot
2\sinh{\frac{\nu_i^{(a)} -\nu_j^{(a)}}{2}}\cdot 2\sinh{\frac{\nu_i^{(a)} + \nu_j^{(a)}}{2}}
\Bigr]^2}
{\prod_{i,j} 
2\cosh{\frac{\mu_i^{(a)} -\nu_j^{(a)}}{2}}\cdot 2\cosh{\frac{\mu_i^{(a)} + \nu_j^{(a)}}{2}}  \cdot
2\cosh{\frac{\mu_i^{(a+1)} -\nu_j^{(a)}}{2}}\cdot 2\cosh{\frac{\mu_i^{(a+1)} + \nu_j^{(a)}}{2}}  
} ,
\end{eqnarray}
where $\mu_i^{(r+1)}=\mu_i^{(1)}$ and\footnote{
We could also include masses and FI-terms.
Then $f^{(a)}(x)$ and $f^{\prime (a)}(x )$ are modified
but the result in this section does not essentially change 
up to this modification.
}
\begin{\eq}
f^{(a)}(x )
=\frac{e^{\frac{ik_a}{2\pi}  x^2} }{\left( 2\cosh{\frac{x}{2}} \right)^{2N_f^{(a)}}} ,\quad
f^{\prime (a)}(x )
=\frac{e^{\frac{ik_a'}{2\pi} x^2} }{\left( 2\cosh{\frac{x}{2}} \right)^{2N_f^{\prime (a)}}} .
\end{\eq}
By similar arguments to sec.~\ref{sec:rhoABJM}, 
we rewrite the partition function as
\begin{eqnarray}
Z_{O(2N+1)\times USp(2N)} (N) \nonumber 
&=& \frac{1}{N!}
\sum_{\sigma\in S_N} (-1)^\sigma 
\int  d^N \mu   \prod_j \rho (\mu_j ,\mu_{\sigma (j)} ) .
\end{eqnarray}
Here the function $\rho (x,y)$ is defined by
\begin{\eqa}
&&\rho_{O(2N+1)\times USp(2N)} (x,y) \NN\\
&&= \frac{\sinh{\frac{x}{2k'}} \sinh{\frac{y}{2k'}} f^{(1)}(x )  }{2\pi k'}
\int \left( \prod_{a=2}^r  \frac{d\mu^{(a)}}{2\pi k'} f^{(a)}(\mu^{(a)} ) \right)
\left( \prod_{b=1}^r  \frac{d\nu^{(b)}}{2\pi k'} \frac{\sinh^2{\frac{\nu^{(b)}}{k'}  }  f^{\prime (b)}(\nu^{(b)} ) }{\cosh^2{\frac{\nu^{(b)}}{2k'}}}\right)   \NN\\
&&  \frac{1}{2\cosh{\left( \frac{x -\nu^{(1)}}{2k'} \right) } \cdot 2\cosh{\left( \frac{x +\nu^{(1)}}{2k'} \right) }    } 
\frac{1}{\prod_{a=1}^{r-1} 2\cosh{\left( \frac{\nu^{(a)} -\mu^{(a+1)}}{2k'} \right) } \cdot 2\cosh{\left( \frac{\nu^{(a)} +\mu^{(a+1)}}{2k'} \right) }     } \NN\\
&&\frac{1}{\prod_{a=2}^{r} 2\cosh{\left( \frac{\nu^{(a)} -\mu^{(a)}}{2k'} \right) } \cdot 2\cosh{\left( \frac{\nu^{(a)} +\mu^{(a)}}{2k'} \right) }     } 
 \frac{1}{ 2\cosh{\left( \frac{\nu^{(r)} -y}{2k'} \right) } \cdot 2\cosh{\left( \frac{\nu^{(r)} +y}{2k'} \right) }    } .
\end{\eqa}
By appropriate similarity transformations, we obtain
\begin{\eqa}
\hat{\rho}_{O(2N+1)\times USp(2N)} (\hat{Q} ,\hat{P}) 
&=& \prod_{a=1}^r f^{(a)}(\hat{Q} ) \frac{1}{2\cosh{\frac{\hat{P}}{2}}} 
\frac{1-\hat{R}}{2} f^{\prime (a)}(\hat{Q} ) 
   \frac{1}{2\cosh{\frac{\hat{P}}{2}}}\frac{1-\hat{R}}{2} \NN\\
&=& \hat{\rho}_{U(N)}(\hat{Q} ,\hat{P})  \frac{1-\hat{R}}{2} ,
\label{eq:rho_general}
\end{\eqa}
where $\hat{\rho}_{U(N)}$ is the density matrix operator associated 
with the non-orientifold theory,
which is obtained by the replacement $O(2N+1),USp(2N) \rightarrow U(N)$ 
in the orientifold theories.
This relation shows that
$\hat{\rho}$ for the orientifold theory
is the projection of the one without the orientifolds.

\subsection{Identification of operators corresponding to orientifolds}
\begin{figure}[t]
\begin{center}
\includegraphics[width=7.4cm]{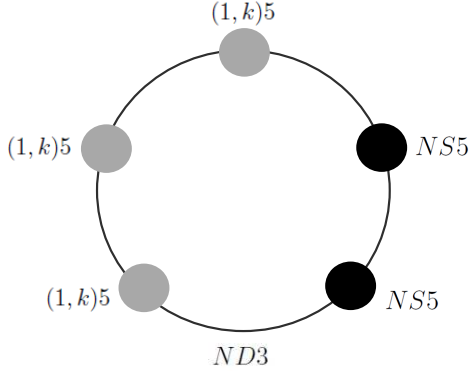}
\end{center}
\caption{
The type IIB brane construction of the $(p,q)$ model for $(p,q)=(2,3)$.
}
\label{fig:brane23}
\end{figure}
Here we identify quantum mechanical operators,
which naturally correspond to the orientifolds\footnote{
$\widetilde{O3}^-$ can be regarded as $O3^-$ plane with a half D3-brane
while $\widetilde{O3}^+$ is perturbatively the same as $O3^+$ plane but different non-perturbatively.
} $\widetilde{O3}^\pm$ in type IIB brane construction.
First, it is known that D5-brane, NS5-brane and $(1,k)$-5 brane 
naturally correspond to\footnote{
We could also consider $(1,\tilde{k})$-5 brane with $\tilde{k}=nk$,
whose corresponding operator is $\hat{\mathcal{O}}_{(1,\tilde{k})5}= e^{\frac{i\tilde{n}Q^2}{2\hbar}} \frac{1}{2\cosh{\frac{P}{2}}} e^{-\frac{i\tilde{n}Q^2}{2\hbar}} = \frac{1}{2\cosh{\frac{P-\tilde{n}Q}{2}}} .$
}
(see e.g.~\cite{Assel:2014awa,Drukker:2015awa}) 
\begin{\eq}
\hat{\mathcal{O}}_{D5}= \frac{1}{2\cosh{\frac{Q}{2}}} ,\quad
\hat{\mathcal{O}}_{NS5}= \frac{1}{2\cosh{\frac{P}{2}}},\quad 
\hat{\mathcal{O}}_{(1,k)5}= e^{\frac{iQ^2}{2\hbar}} \frac{1}{2\cosh{\frac{P}{2}}} e^{-\frac{iQ^2}{2\hbar}} = \frac{1}{2\cosh{\frac{P-Q}{2}}} .
\end{\eq}
This is actually consistent with $\hat{\rho}$ of 
$\mathcal{N}=3$ circular quiver CS theory with $U(N)$ gauge group and
$SL(2,\mathbb{Z})$ symmetry in type IIB string.
For example, 
let us consider the $(p,q)$-model,
whose IIB brane construction consists 
of $p$ NS5-branes and $q$ $(1,k)$-5 branes 
connected by $N$ D3-branes on a circle 
(see fig.~\ref{fig:brane23}).
This theory is $\mathcal{N}=3$ circular quiver superconformal CS theory
with the gauge group $U(N)_k \times U(N)_0^{q-1} \times U(N)_{-k}^{p-1}$
and $\hat{\rho}_{(p,q)}$ associated with this theory is
\begin{\eq} 
\hat{\rho}_{(p,q)} = \hat{\mathcal{O}}_{(1,k)5}^q \hat{\mathcal{O}}_{NS5}^p 
= \left( \frac{1}{2\cosh{\frac{P-Q}{2}}} \right)^q \left( \frac{1}{2\cosh{\frac{P}{2}}} \right)^p ,
\end{\eq}
which is equivalent to $\hat{\rho}$ of the $(p,q)$-model by an appropriate canonical transformation.

\begin{figure}[t]
\begin{center}
\includegraphics[width=7.4cm]{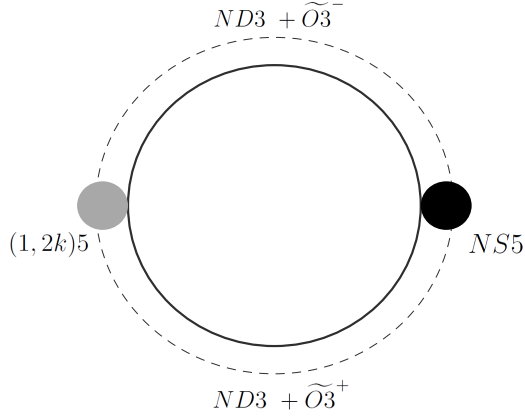}
\end{center}
\caption{
The type IIB brane construction of the $\mathcal{N}=5$ ABJM theory with the gauge group $O(2N+1)_{2k}\times USp(2N)_{-k}$.
}
\label{fig:brane}
\end{figure}
Similarly
let us consider the 
$O(2N+1)_{2k}\times USp(2N)_{-k}$ ABJM theory,
whose brane construction is given by $(\widetilde{O3}^+ -D3)-(1,2k)-(\widetilde{O3}^- -D3) -(NS5)$ on a circle (see fig.~\ref{fig:brane}).
As discussed in sec.~\ref{sec:rhoABJM},
$\hat{\rho}$ for the $\mathcal{N}=5$ ABJM theory is 
\begin{\eq}
\hat{\rho}_{\mathcal{N}=5\rm ABJM}
=e^{\frac{i}{2\hbar}\hat{Q}^2}  \frac{1}{2\cosh{\frac{\hat{P}}{2}}} 
\frac{1-\hat{R}}{2}
  e^{-\frac{i}{2\hbar}\hat{Q}^2} \frac{1}{2\cosh{\frac{\hat{P}}{2}}} \frac{1-\hat{R}}{2}
= \frac{1-\hat{R}}{2} \hat{\mathcal{O}}_{(1,2k)}
 \frac{1-\hat{R}}{2}  \hat{\mathcal{O}}_{NS5}  .
\end{\eq}
If we assume that this can be rewritten as
\begin{\eq}
\hat{\rho}_{\mathcal{N}=5\rm ABJM}
=\hat{\mathcal{O}}_{\widetilde{O3}^+} \hat{\mathcal{O}}_{(1,2k)}
 \hat{\mathcal{O}}_{\widetilde{O3}^-}  \hat{\mathcal{O}}_{NS5}  ,
\end{\eq}
where $\hat{\mathcal{O}}_{\widetilde{O3}^\pm}$ 
corresponds to $\widetilde{O3}^\pm$,
then it is natural to identify\footnote{
As mentioned in remark~2 of sec.~\ref{sec:rhoABJM},
we have multiple representations of $\hat{\rho}$ to describe the same partition function.
Then identifications of $\mathcal{O}_{\tilde{O3}^\pm}$ are more generally
\[
\mathcal{O}_{\widetilde{O3}^-} = 4\sinh^2{\frac{\hat{Q}}{2k}} f_\pm^2 (\hat{Q}) \frac{1\pm\hat{R}}{2}  ,\quad
\mathcal{O}_{\widetilde{O3}^+} = \frac{\sinh^2{\frac{\hat{Q}}{k}}}{\cosh^2{\frac{\hat{Q}}{2k}}} f_\pm^2 (\hat{Q}) \frac{1\pm\hat{R}}{2}  .
\]
}
\begin{\eq}
\hat{\mathcal{O}}_{\widetilde{O3}^-} = \frac{1-\hat{R}}{2}  ,\quad
\hat{\mathcal{O}}_{\widetilde{O3}^+} = \frac{1-\hat{R}}{2}  .
\end{\eq}
This identification is consistent for more general quiver gauge theories
described in sec.~\ref{sec:generalFermi}.

\subsection{Orientifold projection of $(p,q)$-model}
\begin{figure}[t]
\begin{center}
\includegraphics[width=7.4cm]{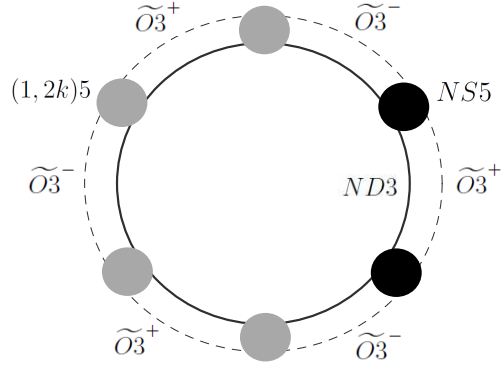}
\end{center}
\caption{
The type IIB brane construction of the orientifold projection
of the $(2,4)$ model.
}
\label{fig:brane24P}
\end{figure}
As an interesting example,
we consider orientifold projection of the $(p,q)$-model 
analyzed well in \cite{Mezei:2013gqa,Grassi:2014vwa,Hatsuda:2014vsa,Moriyama:2014gxa,Moriyama:2014nca,Hatsuda:2015lpa}.
The $(p,q)$-model is the circular quiver theory 
with the gauge group $U(N)_k \times U(N)_0^{q-1}\times U(N)_{-k}\times U(N)_0^{p-1}$,
whose type IIB brane construction is $[(D3)-(NS5)]^p -[(D3)-(1,k)]^q$.
Then let us consider a circular quiver theory with the brane construction 
(see fig.~\ref{fig:brane24P})
\[
[(\widetilde{O3}^- -D3) -(NS5)-(\widetilde{O3}^+ -D3) -(NS5)]^p - [(\widetilde{O3}^- -D3)-(1,2k)-(\widetilde{O3}^+ -D3)-(1,2k)]^q .
\]
Then corresponding $\hat{\rho}$ is
\begin{\eqa} 
\hat{\rho} 
&=& \left( \hat{\mathcal{O}}_{\widetilde{O3}^-}^{(- )} \hat{\mathcal{O}}_{NS5} 
\hat{\mathcal{O}}_{\widetilde{O3}^+}^{(- )} \hat{\mathcal{O}}_{NS5} \right)^p 
\left( \hat{\mathcal{O}}_{\widetilde{O3}^-}^{(- )} 
\hat{\mathcal{O}}_{(1,2k)5} 
\hat{\mathcal{O}}_{\widetilde{O3}^+}^{(- )} 
\hat{\mathcal{O}}_{(1,2k)5} \right)^q \NN\\
&=& \left( \frac{1}{2\cosh{\frac{\hat{P}}{2}}} \right)^{2p}  \left( \frac{1}{2\cosh{\frac{\hat{P}-\hat{Q}}{2}}} \right)^{2q} \frac{1-\hat{R}}{2} 
=\hat{\rho}_{(2p,2q)} \frac{1-\hat{R}}{2} .
\end{\eqa}
This can be understood as the projection of the $(2p,2q)$-model.

We can also consider the orientifold projection of the $(p,q)$-model with odd $p$ and $q$.
For example suppose the brane construction
\[
[(\widetilde{O3}^- -D3) -(1,2k)-(\widetilde{O3}^+ -D3) -(NS5)] - [(\widetilde{O3}^- -D3)-(NS5)-(\widetilde{O3}^+ -D3)-(NS5)]^m ,
\]
which gives
\begin{\eqa} 
\hat{\rho} 
&=&  \hat{\mathcal{O}}_{\widetilde{O3}^-}^{(- )} \hat{\mathcal{O}}_{(1,2k)5}
\hat{ \mathcal{O}}_{\widetilde{O3}^+}^{(- )} \hat{\mathcal{O}}_{NS5} 
\left( \hat{\mathcal{O}}_{\widetilde{O3}^-}^{(- )} \hat{\mathcal{O}}_{NS5} 
\hat{\mathcal{O}}_{\widetilde{O3}^+}^{(- )} 
\hat{\mathcal{O}}_{NS5} \right)^m \NN\\
&= &    \frac{1}{2\cosh{\frac{\hat{P}-\hat{Q}}{2}}}
  \left( \frac{1}{2\cosh{\frac{\hat{P}}{2}}} \right)^{2m+1} \frac{1-\hat{R}}{2}  
=\hat{\rho}_{(1,2m+1)} \frac{1-\hat{R}}{2} .
\end{\eqa}
This is the projection of the $(1,2m+1)$-model.

\subsection{Orientifold projection of orbifold ABJM theory }
\begin{figure}[t]
\begin{center}
\includegraphics[width=7.4cm]{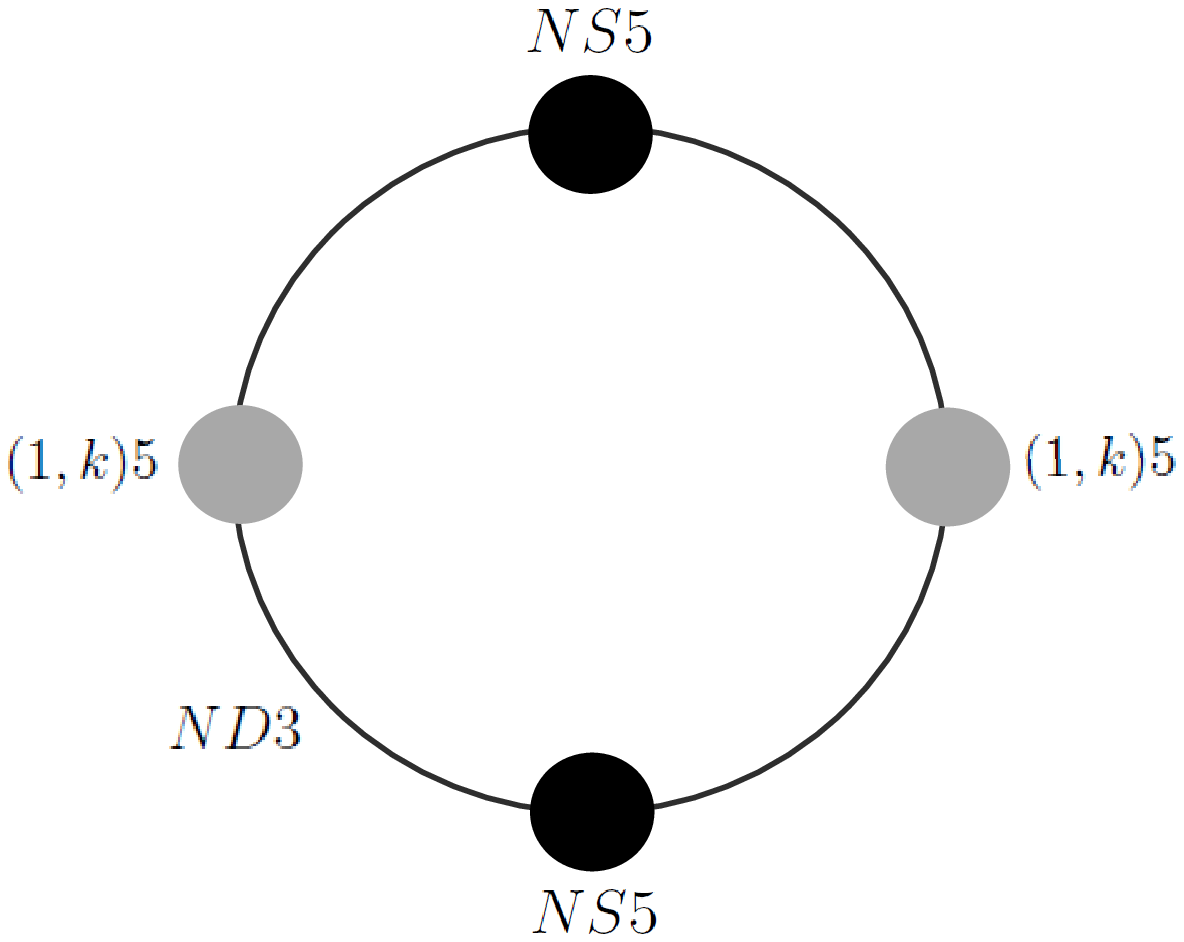}
\includegraphics[width=7.4cm]{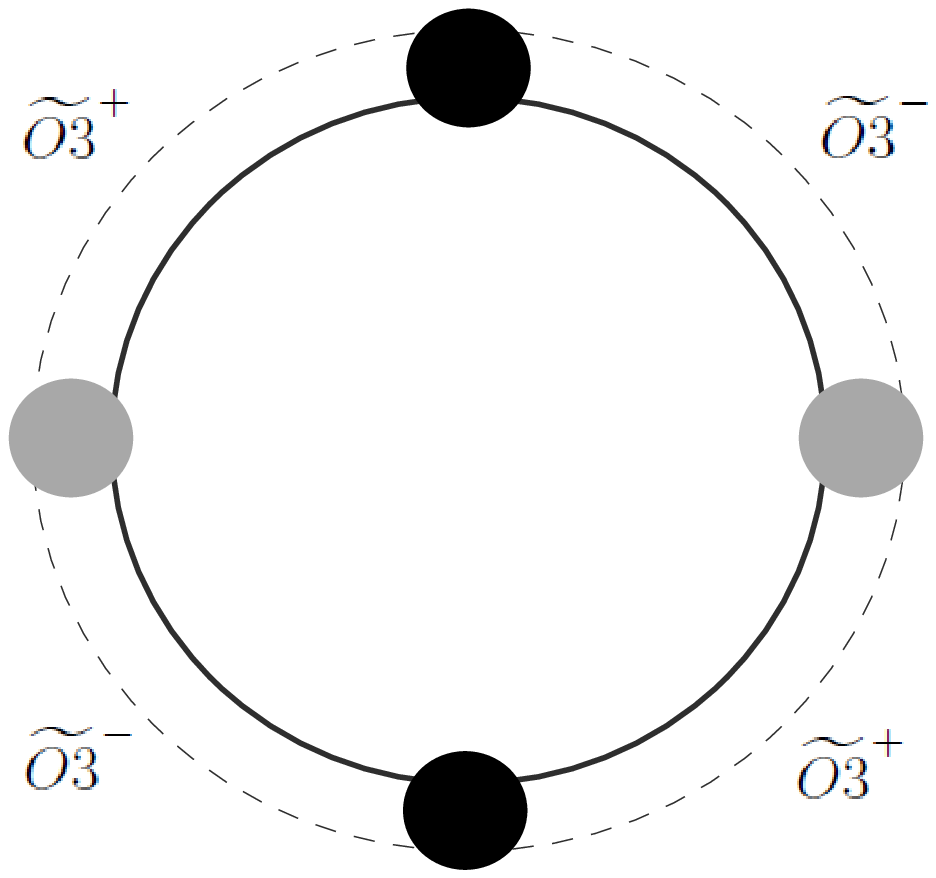}
\end{center}
\caption{
[Left] The type IIB brane construction of the orbifold ABJM theory for $r=2$.
[Right] Its orientifold projection.
}
\label{fig:brane_orbi}
\end{figure}
Next we consider the orientifold projection of the orbifold ABJM theory.
Recalling that the brane construction of the orbifold ABJM theory is $[(D3) -(NS5)-(D3) -(1,k)]^r$,
let us take the following brane construction
(see fig.~\ref{fig:brane_orbi})
\[
[(\widetilde{O3}^- -D3) -(NS5)-(\widetilde{O3}^+ -D3) -(1,2k)]^r ,
\]
which gives the $[O(2N+1)_{2k}\times USp(2N)_{-k}]^r$ circular quiver superconformal CS theory.
Then corresponding $\hat{\rho}$ is
\begin{\eqa} 
\hat{\rho}_{[O(2N+1)_{2k}\times USp(2N)_{-k}]^r} 
&=& \Bigl[ \mathcal{O}_{\widetilde{O3}^-}^{(- )} \mathcal{O}_{NS5} \mathcal{O}_{\widetilde{O3}^+}^{(- )} \mathcal{O}_{(1,2k)5} \Bigr]^r  \NN\\
&= & \left( \frac{1}{2\cosh{\frac{P}{2}}}  \frac{1}{2\cosh{\frac{P-Q}{2}}} \right)^r \frac{1-\hat{R}}{2} 
=\left( \hat{\rho}_{\mathcal{N}=5\rm ABJM} \right)^r ,
\label{eq:orbifoldABJM}
\end{\eqa}
which is the projection of the orbifold ABJM theory.
We can express the grand partition function of 
the orientifold projected orbifold ABJM theory
in terms of the one of the $\mathcal{N}=5$ ABJM theory by using the argument in \cite{Honda:2014ica}.
Namely, when the density matrix operator $\hat{\rho}$ satisfies $\hat{\rho}=(\hat{\rho}')^r$,
the grand partition function becomes
\begin{\eq}
{\rm Det}\left( 1+\rho e^\mu \right) 
=\prod_{j=-\frac{r-1}{2}}^{\frac{r-1}{2}} {\rm Det}\left( 1+\rho' e^{\frac{\mu +2\pi ij}{r}}\right) , 
\end{\eq}
independent of detail form of $\hat{\rho}'$.
Hence, the relation \eqref{eq:orbifoldABJM} immediately leads us to\footnote{
Using the result of \cite{Honda:2014ica},
we can also write ``modified grand potential'' of the orientifold projected orbifold ABJM theory 
in terms of the one of the $\mathcal{N}=5$ ABJM theory.
}
\begin{\eq}
\Xi_{[O(2N+1)_{2k}\times USp(2N)_{-k}]^r} (\mu )
=\prod_{j=-\frac{r-1}{2}}^{\frac{r-1}{2}}
\Xi_{\mathcal{N}=5\rm ABJM} \left( \frac{\mu +2\pi ij}{r} \right) .
\label{Jrelation}
\end{\eq}
Since we already know the exact form of the grand partition function 
for the $O(2N+1)_2 \times USp(2N)_{-1}$ by \eqref{eq:exactk1} and \eqref{eq:k1ABJM},
we can also explicitly write the one of the orientifold projected orbifold ABJM theory with $k=1$ in terms of \eqref{eq:exactk1}.

\subsection{Comments on $O(2N) \times USp(2N)$ type}
\label{sec:O2NUSp}
\begin{table}[t]
\begin{center}
  \begin{tabular}{|c|c |  }
  \hline    Multiplet                                            & One-loop determinant    \\\hline
 $\mathcal{N}=2$  $O(2N)$ vector multiplet  &     
$ \prod_{i<j} \Bigl[ 2\sinh{\frac{\mu_i -\mu_j}{2}}\cdot 2\sinh{\frac{\mu_i + \mu_j}{2}} \Bigr]^2 $   \\\hline
$O(2N)\times USp(2N)$ bi-fund. chiral mult. &
$\left( \prod_{i,j}\Bigl[  2\cosh{\frac{\mu_i -\nu_j}{2}}\cdot 2\cosh{\frac{\mu_i + \nu_j}{2}}
\Bigr]   \right)^{-1}$ \\\hline
  \end{tabular}
\end{center}
\caption{One-loop determinant of each multiplet in the localization of the $O(2N)_{2k} \times USp(2N)_{-k}$ ABJM theory on $S^3$. 
}
\label{tab:1loop_2}
\end{table}
In this section 
we give some comments on partition functions of 
$O(2N)_{2k} \times USp(2N)_{-k}\times \cdots \times O(2N)_{2k} \times USp(2N)_{-k}$ type theories,
recently studied well in \cite{Moriyama:2015asx}.
The $S^3$ partition function of this theory is technically equivalent to 
redefinition of $f^{(a)}(x)$ and $f^{\prime (a)}(x)$ in our analysis presented in sec.~\ref{sec:generalFermi}.
For simplicity, let us consider the $\mathcal{N}=5$ ABJM theory with the gauge group $O(2N)_{2k} \times USp(2N)_{-k}$.
Applying the localization,
the partition function of this theory becomes
\begin{eqnarray}
Z_{O(2N)_{2k}\times USp(2N)_{-k}} \nonumber 
&=& \frac{1}{2^{2N}N!^2} \int \frac{d^N \mu}{(2\pi )^N} \frac{d^N \nu}{(2\pi )^N} 
e^{\frac{ik'}{2\pi}\sum_{j=1}^N (\mu_j^2 -\nu_j^2 )}  \prod_{j=1}^N 4\sinh^2{\nu_j} \NN\\
&&\times
\frac{\prod_{i<j} \Bigl[ 
2\sinh{\frac{\mu_i -\mu_j}{2}}\cdot 2\sinh{\frac{\mu_i + \mu_j}{2}}\cdot
2\sinh{\frac{\nu_i -\nu_j}{2}}\cdot 2\sinh{\frac{\nu_i + \nu_j}{2}}
\Bigr]^2}
{\prod_{i,j}\Bigl[ 
2\cosh{\frac{\mu_i -\nu_j}{2}}\cdot 2\cosh{\frac{\mu_i + \nu_j}{2}} \Bigr]^2} .\NN\\
\end{eqnarray}
By similar arguments to sec.~\ref{sec:generalFermi},
we find
\begin{eqnarray}
&&Z_{O(2N)_{2k}\times USp(2N)_{-k}} \nonumber \\
&&= \frac{1}{2^{2N}N!}
\sum_\sigma (-1)^\sigma \int \frac{d^N \mu}{(2\pi )^N} \frac{d^N \nu}{(2\pi )^N} 
e^{\frac{ik}{2\pi}\sum_{j=1}^N (\mu_j^2 -\nu_j^2 )}  \prod_{j=1}^N 4\sinh^2{\nu_j} \NN\\
&&\ \ \ \times
\prod_j   \frac{1}{2\cosh{\left( \frac{\mu_j -\nu_j}{2} \right) } \cdot
                 2\cosh{\left( \frac{\mu_j +\nu_j}{2} \right) } \cdot
                 2\cosh{\left( \frac{\mu_j -\nu_{\sigma (j)}}{2} \right) } \cdot
                 2\cosh{\left( \frac{\mu_j +\nu_{\sigma (j)}}{2} \right) }  } \NN\\ 
&&= 
\sum_\sigma (-1)^\sigma \int  d^N \mu   
\prod_j \rho_{O(2N)_{2k}\times USp(2N)_{-k}} (\mu_j ,\mu_{\sigma (j)} ) ,
\end{eqnarray}
where
\begin{\eq}
\rho_{O(2N)_{2k}\times USp(2N)_{-k}} (x,y) 
=\frac{1}{2\pi k} \int \frac{d\nu}{2\pi k}    
      \frac{e^{\frac{i}{4\pi k}(x^2 -\nu^2 )} \sinh^2{\frac{\nu}{k}}}
      {2\cosh{\left( \frac{x -\nu}{2k} \right) } \cdot 2\cosh{\left( \frac{x +\nu}{2k} \right) } \cdot
                 2\cosh{\left( \frac{\nu -y}{2k} \right) } \cdot
                  2\cosh{\left( \frac{\nu +y}{2k} \right) }  } .              
\end{\eq}
The quantum mechanical operator $\hat{\rho}_{O(2N)_{2k}\times USp(2N)_{-k}}$
associated with this is
\begin{\eq}
\hat{\rho}_{O(2N)_{2k}\times USp(2N)_{-k}}
= e^{\frac{i}{2\hbar}\hat{Q}^2} f_\pm (\hat{Q}) \frac{1\pm\hat{R}}{2\cosh{\frac{\hat{P}}{2}}} f_\pm (\hat{Q})
  e^{-\frac{i}{2\hbar}\hat{Q}^2}\cdot \sinh^2{\frac{\hat{Q}}{k}} f_\pm (\hat{Q})
\frac{1\pm\hat{R}}{2\cosh{\frac{\hat{P}}{2}}} f_\pm (\hat{Q}) ,
\label{eq:O2NABJM}
\end{\eq}
which is of course the same as the result of \cite{Moriyama:2015asx}.

Next we consider operators corresponding to orientifolds $O3^\pm$.
Let us recall that 
the brane construction of the $O(2N)_{2k'}\times USp(2N)_{-k'}$ ABJM is given by
$(O3^- -D3) -(NS5)- (O3^+ -D3)-(1,2k')$.
The $\hat{\rho}$ for the $O(2N)_{2k}\times USp(2N)_{-k} $ ABJM theory \eqref{eq:O2NABJM} 
can be rewritten as
\begin{\eq}
\hat{\rho}_{O(2N)_{2k}\times USp(2N)_{-k}}
= f_\pm (\hat{Q}) \frac{1\pm\hat{R}}{2} \mathcal{O}_{(1,2k)} f_\pm (\hat{Q})
  \cdot 4\sinh^2{\frac{\hat{Q}}{k}} f_\pm (\hat{Q})
\frac{1\pm\hat{R}}{2}  \mathcal{O}_{NS5} f_\pm (\hat{Q}) .
\end{\eq}
Assuming
$\hat{\rho}_{O(2N)_{2k}\times USp(2N)_{-k}}=\mathcal{O}_{O3^+} \mathcal{O}_{(1,2k)} \mathcal{O}_{O3^-}  \mathcal{O}_{NS5} $,
we arrive at the following identification
\begin{\eq}
\mathcal{O}_{O3^+}^{(\pm )} = 4\sinh^2{\frac{\hat{Q}}{k}} f_\pm^2 (\hat{Q}) \frac{1\pm\hat{R}}{2} ,\quad
\mathcal{O}_{O3^-}^{(\pm )} = f_\pm^2 (\hat{Q}) \frac{1\pm\hat{R}}{2}  .
\end{\eq}

\section{Conclusion and Discussions}
\label{sec:conclusion}
In this paper
we have studied the partition functions of 
the low-energy effective theories of M2-branes,
whose type IIB brane constructions include the orientifolds.
We have mainly focused 
on the circular quiver superconformal CS theory on $S^3$
with the gauge group $O(2N+1)\times USp(2N)\times \cdots \times O(2N+1)\times USp(2N)$,
which is the natural generalization of the $O(2N+1)_{2k} \times USp(2N)_{-k}$ $\mathcal{N}=5$ ABJM theory.
We have found that
the partition function of this type of theory 
have the simple relation \eqref{eq:main}
to the one of the M2-brane theories without the orientifolds
with the gauge group $U(N)\times \cdots \times U(N)$.
By using this relation and the recent results 
in \cite{Grassi:2014uua,Codesido:2014oua},
we have found the exact form \eqref{eq:exactk1}
of the grand partition function of the $O(2N+1)_{2} \times USp(2N)_{-1}$ ABJM theory,
where its SUSY is expected to be enhanced 
to $\mathcal{N}=6$ \cite{Aharony:2008gk}.
As another application,
we discussed that
our result gives the natural physical interpretation of 
the relation \eqref{eq:GHM} 
conjectured by Grassi-Hatsuda-Mari\~no.
We also argued in appendix that
the partition function of $\hat{A}_3$ quiver theory 
has the representation \eqref{eq:D3Fermi} in terms of
an ideal Fermi gas system of $\hat{D}$-type quiver theory and
this leads the relation \eqref{UequalUSp} between the $U(N)$ and $USp(2N)$ SUSY gauge theories.

Our result \eqref{eq:rho_project}, \eqref{eq:rho_general} shows that
the density matrix operator for the orientifold theory is 
the projection of the non-orientifold theory by the operator $(1-\hat{R})/2$.
It is nice if we can understand this relation more precisely.
Our result also implies that
one can systematically study the partition function of the orientifold theory 
by using techniques developed in the studies of the non-orientifold theory.
For instance
the technique introduced in \cite{Okuyama:2015auc} allows us to compute
WKB expansion of ${\rm Tr}(\hat{\rho}^\ell \hat{R})$ systematically\footnote{
We thank Kazumi Okuyama for discussions on this point.
} in terms of information on Wigner transformation of $\hat{\rho}^\ell $.
It is interesting to determine non-perturbative effects in the orientifold theories by such techniques.

Recalling that the $U(N)_k \times U(N)_{-k}$ $\mathcal{N}=6$ ABJM theory 
is described 
by topological string on local $\mathbb{P}^1 \times \mathbb{P}^1$,
this relation would imply that
the $O(2N+1)_{2k}\times USp(2N)_{-k}$ ABJM theory is described by
certain projection in the topological string.
There should be a physical meaning of $(1-\hat{R})/2$ 
in the context of the topological string. 

Although we have found the physical interpretation of one of 
relations conjectured by Grassi-Hatsuda-Mari\~no \cite{Grassi:2014uua},
they also conjectured other relations among the grand partition functions of the ABJ(M) theory with specific values of the parameters:
\begin{\eqa}
&&\Xi_{U(N)_4 \times U(N)_{-4}} (\mu )= \Xi_{U(N+1)_2 \times U(N)_{-2}}^{(+)} (\mu ) ,\quad
\Xi_{U(N+2)_4 \times U(N)_{-4}} (\mu )= \Xi_{U(N+1)_2 \times U(N)_{-2}}^{(-)} (\mu ) ,\NN\\
&& \Xi_{U(N+2)_8 \times U(N)_{-8}} (\mu ) = \Xi_{U(N+2)_4 \times U(N)_{-4}}^{(-)}(\mu )  .   
\end{\eqa}
Although these relations might be accidental coincidences,
it would be illuminating if we can find some physical interpretations.

One of immediate extensions of our analysis is to consider 
the gauge group $O(2N_1 +1)\times USp(2N_2 ) \times \cdots \times O(2N_1 +1)\times USp(2N_2 )$.
Probably this can be done by combining the technique in \cite{Honda:2013pea,Matsumoto:2013nya} with the Cauchy determinant-like formula \eqref{eq:Cauchy}.
If this is the case,
$\hat{\rho}$ for the $O(2N_1 +1)_{2k}\times USp(2N_2 )_{-2k}$ $\mathcal{N}=5$ ABJ(M) theory
would be projection\footnote{
After this paper appeared in arXiv,
this statement is proven in \cite{Moriyama:2016xin}.
} of the one of the $U(N_1 )_{2k} \times U(N_2 )_{-2k}$ $\mathcal{N}=6$ ABJ(M) theory by $(1-\hat{R})/2$.
Another interesting direction is to study other supersymmetric observables such as
supersymmetric Wilson loops.
Then the techniques established in \cite{Hatsuda:2013yua} would be efficient.

\subsection*{Acknowledgment}
We are grateful to Kazumi Okuyama
for his early collaboration and many valuable discussions. 
We thank Benjamin Assel and Sanefumi Moriyama for helpful discussions.

\appendix
\section{An exact relation between $USp(2N)$ and $U(N)$ gauge theories}
\label{sec:A3D3}
In this appendix
we show the exact relation \eqref{UequalUSp} between the SUSY gauge theories with $U(N)$ and $USp(2N)$ gauge groups.
The $U(N)$ gauge theory, which we consider here, is
$\mathcal{N}=4$ vector multiplet with one adjoint hyper multiplet and $N_f$ fundamental hyper multiplets,
whose partition function is described by so-called $N_f$-matrix model \cite{Grassi:2014vwa}:
\begin{eqnarray}
Z_{U+adj.}(N,N_f ) 
= \frac{1}{N!} \int \frac{d^N \mu}{(2\pi )^N}
  \prod_{j=1}^N \frac{1}{\left( 2\cosh{\frac{\mu_j}{2}} \right)^{N_f}}
\prod_{i<j}  \tanh^2{\frac{\mu_i -\mu_j}{2}}  .
\label{eq:Nf_matrix}
\end{eqnarray}
This matrix model has been studied well 
in \cite{Mezei:2013gqa,Grassi:2014vwa,Moriyama:2014gxa,Moriyama:2014nca,Hatsuda:2015lpa}.
The $USp(2N)$ gauge theory is
$\mathcal{N}=4$ vector multiplet with one anti-symmetric hyper multiplet and $N_f$-fundamental hyper multiples and
its partition function is
\begin{eqnarray}
Z_{USp+A}(N, N_f ) 
= \frac{1}{2^{2N}N!} \int \frac{d^N \mu}{(2\pi )^N}
  \prod_{j=1}^N \frac{4\sinh^2{\mu_j}}{\left( 4\cosh^2{\frac{\mu_j}{2}} \right)^{N_f}}
\prod_{i<j} \Biggl[ \frac{ \sinh{\frac{\mu_i -\mu_j}{2}}\cdot \sinh{\frac{\mu_i + \mu_j}{2}} }
   { \cosh{\frac{\mu_i -\mu_j}{2}}\cdot \cosh{\frac{\mu_i + \mu_j}{2}} } \Biggr]^2 ,
\end{eqnarray}
which has been analyzed  
in \cite{Mezei:2013gqa,Assel:2015hsa,Okuyama:2015auc}.
Regarding these theories,
the work \cite{Okuyama:2015auc} has proposed
the following equivalence\footnote{
One can also compare this with the $O(2N+1)$ gauge theory with $\mathcal{N}=4$ vector multiplet, 
one anti-symmetric hyper multiplet and $N_f$-fundamental hyper multiples ($O(2N+1)+A$ theory).
Then because of $Z_{O(2N+1)+A}(N, N_f  )=Z_{USp+A}(N, N_f -2 )$,
the relation \eqref{UequalUSp} also indicates $Z_{U(N)+adj.} (N,N_f =4 ) =Z_{O(2N+1)+A}(N,N_f =5 )$.
}
\[
Z_{U(N)+adj.} (N,N_f =4 ) =Z_{USp(2N)+A}(N,N_f =3 ) .
\]
This relation is expected from 3d mirror symmetry \cite{Intriligator:1996ex,deBoer:1996mp}.
It is known that 
the $U(N)$ and $USp(2N)$ theories are equivalent to $\hat{A}_{N_f -1}$ and $\hat{D}_{N_f }$ quiver theories without CS levels,
where only one of the vector multiples is coupled to one fundamental hyper multiplet, respectively.
Since $\hat{A}_3 =\hat{D}_3$, the equation \eqref{UequalUSp} should hold.
In this section we explicitly prove this relation by using the technique in \cite{Assel:2015hsa,Moriyama:2015jsa}.

\subsection{$\hat{A}_3 =\hat{D}_3$}
\label{sec:A3D3proof}
Here we show that
partition function of $\hat{A}_3$ quiver theory has a representation in terms of
an ideal Fermi gas system of $\hat{D}$-type quiver theory.
Although this may be already proven in \cite{Assel:2015hsa,Moriyama:2015jsa},
it is unclear to us whether their derivation includes our analysis in this section or not
and therefore we explicitly prove it.

First we precisely explain what we would like to prove.
Suppose the SUSY CS theory with $\hat{A}_n$ quiver,
namely the circular quiver with the gauge group $U(N)_{k_1} \times \cdots U(N)_{k_{n+1}}$,
which is coupled to $N_f^{(a)}$ fundamental hyper multiplets.
The partition function of the $\hat{A}_n$ quiver theory can be denoted by \cite{Kapustin:2009kz}
\begin{eqnarray}
Z_{\hat{A}_n} \nonumber 
&=& \frac{1}{N!^{n+1}} \int \prod_{a=1}^{n+1} \frac{d^N \mu^{(a)}}{(2\pi )^N}  
   \prod_{j=1}^N  f^{(a)}(\mu_j^{(a)} ) 
\frac{\prod_{i<j} \Bigl[ 
2\sinh{\frac{\mu_i^{(a)} -\mu_j^{(a)}}{2}}\cdot 2\sinh{\frac{\mu_i^{(a)} + \mu_j^{(a)}}{2}}\Bigr]^2}
{\prod_{i,j}  2\cosh{\frac{\mu_i^{(a)} -\mu_j^{(a+1)}}{2}}} ,
\end{eqnarray}
where $\mu_j^{(n+2)}=\mu_j^{(1)}$ and
\begin{\eq}
f^{(a)}(x )
=\frac{e^{\frac{ik_a}{2\pi}  x^2} }{\left( 2\cosh{\frac{x}{2}} \right)^{2N_f^{(a)}}} .
\end{\eq}
It is known that
one can rewrite the partition function of the $\hat{A}_n$ theory 
as \cite{Marino:2011eh}
\begin{\eq}
Z_{\hat{A}_n}(N) =\sum_{\sigma\in S_N} (-1)^\sigma \int d^N x \prod_{j=1}^N \rho_{\hat{A}_n} (x_j ,x_{\sigma (j)}) ,
\end{\eq}
where $\rho_{\hat{A}_n}(x,y)$ is the density matrix of the ideal Fermi gas system
associated with the quantum mechanical operator
\begin{\eq}
\hat{\rho}_{\hat{A}_n} (\hat{Q} ,\hat{P}) 
= \prod_{a=1}^{n+1} f^{(a)}(\hat{Q} ) \frac{1}{2\cosh{\frac{\hat{P}}{2}}}  .
\end{\eq}

\begin{figure}[t]
\begin{center}
\includegraphics[width=7.4cm]{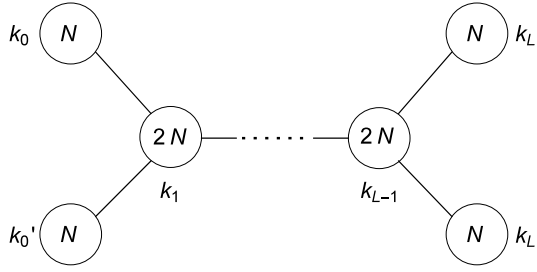}
\end{center}
\caption{
The $\hat{D}_{L+2}$ quiver diagram.
}
\label{fig:Dquiver}
\end{figure}
Next let us consider the $\hat{D}_{L+2}$ quiver CS theory
with the gauge group $U(N)_{k_0} \times U(N)_{k_0'}\times U(2N)_{k_1} \times \cdots U(2N)_{k_{L-1}}\times U(N)_{k_L} \times U(N)_{k_L'}$
(see fig.~\ref{fig:Dquiver}) .
The partition function of this theory 
is given by \cite{Kapustin:2009kz}
\begin{eqnarray}
Z_{\hat{D}_{L+2}} 
&=& \frac{1}{N!^4 (2N!)^{L-1}} \int  \frac{d^N \mu^{(0)}}{(2\pi )^N}\frac{d^N \mu^{\prime (0)}}{(2\pi )^N} 
\frac{d^N \mu^{(L)}}{(2\pi )^N}\frac{d^N \mu^{\prime (L)}}{(2\pi )^N}
 \prod_{a=1}^{L-1} \frac{d^{2N} \mu^{(a)}}{(2\pi )^{2N}}
 \prod_{a=1}^{L-1} \prod_{J=1}^{2N} F^{(a)}(\mu_J^{(a)} )  \NN\\
&&   \prod_{j=1}^N  F^{(0)}(\mu_j^{(0)} ) F^{\prime (0)}(\mu_j^{\prime (0)} ) 
  F^{(L)}(\mu_j^{(L)} ) F^{\prime (L)}(\mu_j^{\prime (L)} )  
 \frac{\prod_{a=1}^{L-1} \prod_{I\neq J}  2\sinh{\frac{\mu_I^{(a)} -\mu_J^{(a)}}{2}}  }
{\prod_{a=1}^{L-2} \prod_{I,J} 2\cosh{\frac{\mu_I^{(a)} -\mu_J^{(a+1)}}{2}} }     \NN\\
&& \frac{\prod_{i\neq j} 2\sinh{\frac{\mu_i^{(0)} -\mu_j^{(0)}}{2}}  \cdot 2\sinh{\frac{\mu_i^{\prime (0)} -\mu_j^{\prime (0)}}{2}}  \cdot
2\sinh{\frac{\mu_i^{(L)} -\mu_j^{(L)}}{2}}  \cdot 2\sinh{\frac{\mu_i^{\prime (L)} -\mu_j^{\prime (L)}}{2}} }
{\prod_{i=1}^N \prod_{J=1}^{2N}  2\cosh{\frac{\mu_i^{(0)} -\mu_J^{(1)}}{2}} \cdot  2\cosh{\frac{\mu_i^{\prime (0)} -\mu_J^{(1)}}{2}} \cdot
  2\cosh{\frac{\mu_i^{(L)} -\mu_J^{(L-1)}}{2}} \cdot  2\cosh{\frac{\mu_i^{\prime (L)} -\mu_J^{(L-1)}}{2}} } , \NN\\
\end{eqnarray}
where
\begin{\eq}
F^{(a)}(x )
=\frac{e^{\frac{ik_a}{2\pi}  x^2} }{\left( 2\cosh{\frac{x}{2}} \right)^{2N_f^{(a)}}} ,\quad
F^{\prime (a)}(x )
=\frac{e^{\frac{ik_a'}{2\pi}  x^2} }{\left( 2\cosh{\frac{x}{2}} \right)^{2N_f^{\prime (a)}}} .
\end{\eq}
It is also known that
the partition function of the $\hat{D}_{L+2}$ quiver theory 
is described by an ideal Fermi gas system \cite{Assel:2015hsa,Moriyama:2015jsa}:
\begin{\eq}
Z_{\hat{D}_{L+2}} 
= \frac{1}{N!} \sum_{\sigma\in S_{N}} (-1)^\sigma \int  \frac{d^N x}{(2\pi )^{N}}
\prod_{j=1}^N  \rho_{\hat{D}_{L+2}}^{(\pm )} (x_j ,x_{\sigma (j)}) ,
\end{\eq}
where
\begin{\eqa}
&&\hat{\rho}_{\hat{D}_{L+2}}\NN\\
&&=
 \left( F^{(0)}(\hat{Q})\tanh{\frac{\hat{P}}{2}}F^{\prime (0)}(\hat{Q}) +F^{\prime (0)}(\hat{Q})\tanh{\frac{\hat{P}}{2}}F^{(0)}(\hat{Q}) \right)
       \frac{1}{2\cosh{\frac{\hat{P}}{2}}} \left( \prod_{a=1}^{L-1} F^{(a)}(\hat{Q})  \frac{1}{2\cosh{\frac{\hat{P}}{2}}} \right) \NN\\
&& \left( F^{(L)}(\hat{Q})\tanh{\frac{\hat{P}}{2}}F^{\prime (L)}(\hat{Q}) +F^{\prime (L)}(\hat{Q})\tanh{\frac{\hat{P}}{2}}F^{(L)}(\hat{Q}) \right)
\frac{1}{2\cosh{\frac{\hat{P}}{2}}} \left( \prod_{a=1}^{L-1} F^{(L-a)}(\hat{Q})  \frac{1}{2\cosh{\frac{\hat{P}}{2}}} \right) . \NN\\
\end{\eqa}
In this section we prove 
\begin{\eq}
Z_{\hat{A}_3} 
= \frac{1}{N!} \sum_{\sigma\in S_{N}} (-1)^\sigma \int  \frac{d^N x}{(2\pi )^{N}}
\prod_{j=1}^N \Biggl[ \lim_{L\rightarrow 1} \rho_{\hat{D}_{L+2}}^{(\pm )} (x_j ,x_{\sigma (j)}) \Biggr] ,
\label{eq:D3Fermi}
\end{\eq}
where
\begin{\eqa}
\lim_{L\rightarrow 1} \hat{\rho}_{\hat{D}_{L+2}}
&=& 
\left( F^{(0)}(\hat{Q})\tanh{\frac{\hat{P}}{2}}F^{\prime (0)}(\hat{Q}) +F^{\prime (0)}(\hat{Q})\tanh{\frac{\hat{P}}{2}}F^{(0)}(\hat{Q}) \right)
       \frac{1}{2\cosh{\frac{\hat{P}}{2}}}  \NN\\
&& \left( F^{(1)}(\hat{Q})\tanh{\frac{\hat{P}}{2}}F^{\prime (1)}(\hat{Q}) +F^{\prime (1)}(\hat{Q})\tanh{\frac{\hat{P}}{2}}F^{(1)}(\hat{Q}) \right)
\frac{1}{2\cosh{\frac{\hat{P}}{2}}}  . 
\end{\eqa}
As mentioned above,
this may be already proven in \cite{Assel:2015hsa,Moriyama:2015jsa}.
However their derivation apparently seems to take $L\geq 2$,
where at least one $U(2N)$ node is present, and
it is unclear to us whether their derivation includes $\hat{D}_3$ ($L=1$) case or not.
Therefore we explicitly prove this relation.

Now let us consider the $A_3$ quiver theory:
\begin{eqnarray}
Z_{\hat{A}_3} \nonumber 
&=& \frac{1}{N!^4} \int \prod_{a=1}^4 \frac{d^N \mu^{(a)}}{(2\pi )^N}  
   \prod_{j=1}^N  f^{(a)}(\mu_j^{(a)} ) 
\frac{\prod_{i<j} \Bigl[ 
2\sinh{\frac{\mu_i^{(a)} -\mu_j^{(a)}}{2}}\cdot 2\sinh{\frac{\mu_i^{(a)} + \mu_j^{(a)}}{2}}\Bigr]^2}
{\prod_{i,j}  2\cosh{\frac{\mu_i^{(a)} -\mu_j^{(a+1)}}{2}}}  .
\end{eqnarray}
Let us redefine the variables as in $\hat{D}_3$-quiver language:
\begin{\eqa}
&&\mu_j^{(1)} = x_j ,\quad \mu_j^{(2)} = y_{N+j} ,\quad \mu_j^{(3)} = x_{N+j} ,\quad \mu_j^{(4)} = y_j ,\NN\\
&& F^{(0)}(x) = f^{(1)}(x),\quad F^{\prime (0)}(x) = f^{(3)}(x),\quad F^{(1)}(x) = f^{(4)}(x),\quad F^{\prime (1)}(x) = f^{(2)}(x) . \NN\\
\end{\eqa}
Then the partition function becomes
\begin{\eqa}
Z_{\hat{A}_3} 
&=& \frac{1}{N!^4} \int  \frac{d^{2N} x}{(2\pi )^{2N}} \frac{d^{2N} y}{(2\pi )^{2N}}
  \prod_{j=1}^N  F^{(0)}(x_j ) F^{\prime (0)}(x_{N+j} )   F^{(1)}(y_j  ) F^{\prime (1)}(y_{N+j} ) \NN\\
&& \frac{\Bigl[ \prod_{i<j} 
 2\sinh{\frac{x_i -x_j}{2}} \cdot 2\sinh{\frac{x_{N+i} -x_{N+j}}{2}} \cdot 2\sinh{\frac{y_i -y_j}{2}} \cdot 2\sinh{\frac{y_{N+i} -y_{N+j}}{2}} \Bigr]^2}
            {\prod_{I,J}  2\cosh{\frac{x_I -y_J}{2}} } , 
\end{\eqa}
where $I,J =1,\cdots ,2N$. 
By inserting
\begin{\eq}
1 = \frac{\prod_{i,j} 2\sinh{\frac{x_i -x_{N+j}}{2}} \cdot 2\sinh{\frac{y_i -y_{N+j}}{2}} }
             {\prod_{i,j} 2\sinh{\frac{x_i -x_{N+j}}{2}} \cdot 2\sinh{\frac{y_i -y_{N+j}}{2}}} ,
\end{\eq} 
to the integrand and using the Cauchy determinant formula,
we find
\begin{\eqa}
Z_{\hat{A}_3} 
&=& \frac{1}{N!^2} \sum_{\sigma\in S_{2N}} (-1)^\sigma \int  \frac{d^{2N} x}{(2\pi )^{2N}} \frac{d^{2N} y}{(2\pi )^{2N}}
\prod_{j=1}^N  F^{(0)}(x_j ) F^{\prime (0)}(x_{N+j} )   F^{(1)}(y_j  ) F^{\prime (1)}(y_{N+j} ) \NN\\
&& \times\frac{1}{ \prod_{j=1}^N  2\sinh{\frac{x_j -x_{N+j}}{2}} \cdot 2\sinh{\frac{y_j -y_{N+j}}{2}} }  
 \frac{1}{ \prod_{J=1}^{2N}  2\cosh{\frac{x_J -y_{\sigma (J)}}{2}} } . 
\end{\eqa}
Below in this subsection
we just repeat the argument of \cite{Assel:2015hsa}. 
According to \cite{Assel:2015hsa}, we introduce
\begin{\eq}
R(j)=N+j,\quad R(N+j) =j .
\end{\eq}
Now we would like to rewrite the integral in terms of a kernel acting on 
set of $N$ eigenvalues $\mathcal{K}(\sigma )$ among $x_J$'s, which is dependent on the permutation $\sigma$.
More precisely, we take $\mathcal{K}(\sigma )$ 
such that $R\tau^{-1}R\tau (j) \in\mathcal{K}(\sigma ) $ for given $j\in \mathcal{K}(\sigma )$.
Then we rewrite the partition function as
\begin{\eqa}
Z_{\hat{A}_3} 
&=& \frac{1}{N!^2} \sum_{\sigma\in S_{2N}} (-1)^\sigma \int  \frac{d^{2N} x}{(2\pi )^{2N}} \frac{d^{2N} y}{(2\pi )^{2N}} 
\prod_{j=1}^N  F^{(0)}(x_j ) F^{\prime (0)}(x_{N+j} )   F^{(1)}(y_j  ) F^{\prime (1)}(y_{N+j} ) \NN\\
&& \prod_{j\in\mathcal{K}(\sigma )} \frac{1}{  2\cosh{\frac{x_j -y_{\sigma (j)}}{2}}}
\frac{(-1)^{s(\sigma (j))}}{2\sinh{\frac{y_{\sigma (j)} -y_{R\sigma (j)}}{2}}}
 \frac{1}{ 2\cosh{\frac{y_{R\sigma (j )} -x_{\sigma^{-1}R\sigma (j)}}{2}}  }  
\frac{(-1)^{s(\sigma^{-1}R\sigma (j))}}{2\sinh{\frac{x_{\sigma^{-1}R\sigma (j)} -x_{R\sigma^{-1}R\sigma (j)} }{2}}} ,\NN\\
\label{eq:exp1}
\end{\eqa}
where
\begin{\eq}
s(j) = \left\{ \begin{matrix} 
0 & {\rm for}& j=1,\cdots ,N \cr 
1 & {\rm for}&  j=N+1,\cdots ,2N \cr 
\end{matrix} \right. .
\end{\eq}
Note that we can also write this as
\begin{\eqa}
Z_{\hat{A}_3} 
&=& \frac{1}{N!^2} \sum_{\sigma\in S_{2N}} (-1)^{R\sigma} \int  \frac{d^{2N} x}{(2\pi )^{2N}} \frac{d^{2N} y}{(2\pi )^{2N}} 
\prod_{j=1}^N  F^{(0)}(x_j ) F^{\prime (0)}(x_{N+j} )   F^{(1)}(y_j  ) F^{\prime (1)}(y_{N+j} ) \NN\\
&& \prod_{j\in\mathcal{K}(\sigma )} \frac{1}{  2\cosh{\frac{x_j -y_{R\sigma (j)}}{2}}}
\frac{(-1)^{s(R\sigma (j))}}{2\sinh{\frac{y_{R\sigma (j)} -y_{\sigma (j)}}{2}}}
 \frac{1}{ 2\cosh{\frac{y_{\sigma (j )} -x_{\sigma^{-1}R\sigma (j)}}{2}}  }  
\frac{(-1)^{s(\sigma^{-1}R\sigma (j))}}{2\sinh{\frac{x_{\sigma^{-1}R\sigma (j)} -x_{R\sigma^{-1}R\sigma (j)} }{2}}} .\NN\\
\label{eq:exp2}
\end{\eqa}
Averaging over these,
we obtain
\begin{\eqa}
Z_{\hat{A}_3} 
&=& \frac{1}{2^N N!^2} \sum_{\sigma\in S_{2N}} (-1)^\sigma \int  \frac{d^{2N} x}{(2\pi )^{2N}} \frac{d^{2N} y}{(2\pi )^{2N}} 
\prod_{j=1}^N  F^{(0)}(x_j ) F^{\prime (0)}(x_{N+j} )   F^{(1)}(y_j  ) F^{\prime (1)}(y_{N+j} ) \NN\\
&&\prod_{j\in\mathcal{K}(\sigma )} (-1)^{s(\sigma (j)+s(j) +1}
\Biggl[ 
\frac{1}{  2\cosh{\frac{x_j -y_{\sigma (j)}}{2}}}
\frac{1}{2\sinh{\frac{y_{\sigma (j)} -y_{R\sigma (j)}}{2}}} \frac{1}{ 2\cosh{\frac{y_{R\sigma (j )} -x_{\sigma^{-1}R\sigma (j)}}{2}}  }  \NN\\
&&+\frac{1}{  2\cosh{\frac{x_j -y_{R\sigma (j)}}{2}}}
\frac{1}{2\sinh{\frac{y_{R\sigma (j)} -y_{\sigma (j)}}{2}}} \frac{1}{ 2\cosh{\frac{y_{\sigma (j )} -x_{\sigma^{-1}R\sigma (j)}}{2}}  }   
\Biggr] \frac{1}{2\sinh{\frac{x_{\sigma^{-1}R\sigma (j)} -x_{R\sigma^{-1}R\sigma (j)} }{2}}} \NN\\
&=& \frac{1}{2^{2N} N!^2} \sum_{\sigma\in S_{2N}} (-1)^\sigma \int  \frac{d^N x}{(2\pi )^{N}}
\prod_{j\in\mathcal{K}(\sigma )} (-1)^{s(\sigma (j)+s(j) } \rho (x_j ,x_{R\sigma^{-1}R\sigma (j)}) \NN\\
&=& \frac{1}{N!} \sum_{\sigma\in S_{N}} (-1)^\sigma \int  \frac{d^N x}{(2\pi )^{N}}
\prod_{j=1}^N  \rho_{\hat{D}_3}^{(\pm )} (x_j ,x_{\sigma (j)}) ,
\end{\eqa}
where
\begin{\eqa}
&&\rho_{\hat{D}_3} (x_1 ,x_2 )\NN\\
&&= -2 \int \frac{dydy'dx' }{(2\pi )^3}
 \frac{1}{  2\cosh{\frac{x-y}{2}}} 
\left( F^{(0)}(y) \frac{1}{2\sinh{\frac{y -y'}{2}}} F^{\prime (0)}(y' ) +F^{\prime (0)}(y ) \frac{1}{2\sinh{\frac{y -y'}{2}}}  F^{(0)}(y') \right) \NN\\
&& \frac{1}{ 2\cosh{\frac{y' -x'}{2}}  }
\left( F^{(1)}(x') \frac{1}{2\sinh{\frac{x' -x}{2}}} F^{\prime (1)}(x ) +F^{\prime (1)}(x' ) \frac{1}{2\sinh{\frac{x' -x}{2}}}  F^{(1)}(x) \right)  .
\end{\eqa}
Hence corresponding operator $\hat{\rho}_{\hat{D}_3}$ is 
\begin{\eqa *}
\hat{\rho}_{\hat{D}_3} 
&=& 
\left( F^{(0)}(\hat{Q})\tanh{\frac{\hat{P}}{2}}F^{\prime (0)}(\hat{Q}) +F^{\prime (0)}(\hat{Q})\tanh{\frac{\hat{P}}{2}}F^{(0)}(\hat{Q}) \right)
       \frac{1}{2\cosh{\frac{\hat{P}}{2}}}  \NN\\
&& \left( F^{(1)}(\hat{Q})\tanh{\frac{\hat{P}}{2}}F^{\prime (1)}(\hat{Q}) +F^{\prime (1)}(\hat{Q})\tanh{\frac{\hat{P}}{2}}F^{(1)}(\hat{Q}) \right)
\frac{1}{2\cosh{\frac{\hat{P}}{2}}}  \NN\\
&=&\lim_{L\rightarrow 1} \hat{\rho}_{\hat{D}_{L+2}} .
\end{\eqa *}

\subsection{$\hat{A}_n \rightarrow U(N)+adj.$}
\label{sec:mirrorA}
Suppose 
the $\hat{A}_n$ quiver theories without CS terms,
where only one of the $U(N)$ vector multiples is coupled to one fundamental hyper multiplet.
This theory is related to 
the $U(N)$ gauge theory
with $\mathcal{N}=4$ vector multiplet, one adjoint hyper multiplet 
and $n+1$ fundamental hyper multiplets.
We can easily show this for the partition functions \cite{Kapustin:2010xq,Marino:2011eh,Drukker:2015awa}.
To be self contained,
here we repeat its derivation.
The density matrix operator $\hat{\rho}$ of the $\hat{A}_n$ theory is
\begin{\eq}
\rho_{\hat{A}_n} 
= \frac{1}{2\cosh{\frac{\hat{Q}}{2}}} \frac{1}{2\cosh^{n+1}{\frac{\hat{P}}{2}}}
\end{\eq}
By the canonical transformation $(Q,P)\rightarrow (Q,-P)$, we get
\begin{\eq}
\rho_{\hat{A}_n} = \frac{1}{2\cosh{\frac{\hat{P}}{2}}} \frac{1}{2\cosh^{n+1}{\frac{\hat{Q}}{2}}} .
\end{\eq}
This $\hat{\rho}$ gives the $N_f$ matrix model \eqref{eq:Nf_matrix} 
with $N_f =n+1$.

\subsection{$\hat{D}_n \rightarrow USp +A$}
\label{sec:mirrorUSp}
We also review 
the proof of the 3d mirror symmetry
between the partition functions on $S^3$ of 
the $\hat{D}_n$ quiver and $USp+A$ theories.
The gauge group of the $\hat{D}_n$ quiver theory consists of
four $U(N)$ nodes and $(n-3)$ $U(2N)$ nodes,
where one of $U(N)$ nodes associates one fundamental hypermultiplet.
The partition function of this theory is given by
\begin{eqnarray}
Z_{\hat{D}_n} 
&=& \frac{1}{N!^2 (2N!)^{n-3}} \int \frac{d^N \mu}{(2\pi )^N} \frac{d^N \mu'}{(2\pi )^N}  \frac{d^N \nu}{(2\pi )^N} \frac{d^N \nu'}{(2\pi )^N}
   \frac{d^{2N} \lambda^{(1)}}{(2\pi )^{2N}} \cdots \frac{d^{2N} \lambda^{(n-3)}}{(2\pi )^{2N}} \NN\\
&&  \frac{ \prod_{i\neq j} 2\sinh{\frac{\mu_i -\mu_j}{2}}\cdot 2\sinh{\frac{\mu_i' - \mu_j'}{2}}  }
   {\prod_j 2\cosh{\frac{\mu_j}{2}} \prod_{i,J}   2\cosh{\frac{\mu_i -\lambda^{(1)}_J}{2}}\cdot  2\cosh{\frac{\mu'_i -\lambda^{(1)}_J}{2}} }  \NN\\
&& \prod_{\alpha =1}^{n-4}\Biggl[ \frac{  \prod_{I\neq J}  2\sinh{\frac{\lambda_I^{(\alpha )} -\lambda_J^{(\alpha )}}{2}} \cdot
    2\sinh{\frac{\lambda_I^{(\alpha +1)} -\lambda_J^{(\alpha +1)}}{2}} }
   { \prod_{I,J} 2\cosh{\frac{\lambda_I^{(\alpha )} -\lambda_J^{(\alpha +1)}}{2}} }  \Biggr] 
  \frac{ \prod_{i\neq j} 2\sinh{\frac{\nu_i -\nu_j}{2}}\cdot 2\sinh{\frac{\nu_i' - \nu_j'}{2}}  }
   {\prod_{i,J}   2\cosh{\frac{\nu_i -\lambda^{(n-3)}_J}{2}}\cdot  2\cosh{\frac{\nu'_i -\lambda^{(n-3)}_J}{2}} } . \NN\\
\end{eqnarray}
Corresponding $\hat{\rho}$ is
\begin{\eqa}
\hat{\rho}_{\hat{D}_n}
&=& 
 \left\{ \frac{1}{2\cosh{\frac{\hat{Q}}{2}}} , \tanh{\frac{\hat{P}}{2}}  \right\}
   \left( \frac{1}{2\cosh^{n-2}{\frac{\hat{P}}{2}}}  \right)^{n-2}
    \tanh{\frac{\hat{P}}{2}} 
   \left( \frac{1}{2\cosh^{n-2}{\frac{\hat{P}}{2}}}  \right)^{n-2} .
\label{eq:Dquiver}
\end{\eqa}
By using
\begin{\eq}
 \left\{ \frac{1}{\cosh{\frac{\hat{Q}}{2}}} , \tanh{\frac{\hat{P}}{2}}  \right\}
= \frac{2}{\cosh{\frac{\hat{P}}{2}}} 
 \left( \sinh{\frac{\hat{P}}{2}}\frac{1}{\cosh{\frac{\hat{Q}}{2}}} \cosh{\frac{\hat{P}}{2}} 
         +\cosh{\frac{\hat{P}}{2}}\frac{1}{\cosh{\frac{\hat{Q}}{2}}}\sinh{\frac{\hat{P}}{2}} \right)
\frac{1}{\cosh{\frac{\hat{P}}{2}}} ,\NN\\
\end{\eq}
we find
\begin{\eqa}
\hat{\rho}_{\hat{D}_n}
&=& \frac{1}{2\cosh{\frac{\hat{P}}{2}}}
 \left( 2\sinh{\frac{\hat{P}}{2}}\frac{1}{2\cosh{\frac{\hat{Q}}{2}}} 2\cosh{\frac{\hat{P}}{2}} 
         +2\cosh{\frac{\hat{P}}{2}}\frac{1}{2\cosh{\frac{\hat{Q}}{2}}}
         2\sinh{\frac{\hat{P}}{2}} \right) \NN\\
&&   \left( \frac{1}{2\cosh^{n-2}{\frac{\hat{P}}{2}}}  \right)^{n-1}
  \tanh{\frac{\hat{P}}{2}} 
   \left( \frac{1}{2\cosh^{n-2}{\frac{\hat{P}}{2}}}  \right)^{n-2} .
\end{\eqa}
Then the similarity transformation
\begin{\eq}
\hat{\rho}_{\hat{D}_n}
\rightarrow 
2\cosh{\frac{\hat{P}}{2}} \cdot \hat{\rho} \cdot \frac{1}{2\cosh{\frac{\hat{P}}{2}}} ,
\end{\eq}
leads us to
\begin{\eq}
\hat{\rho}_{\hat{D}_n}
= 
\left(2 \sinh{\frac{\hat{P}}{2}}\frac{1}{2\cosh{\frac{\hat{Q}}{2}}} 2\cosh{\frac{\hat{P}}{2}} 
      +2\cosh{\frac{\hat{P}}{2}}\frac{1}{2\cosh{\frac{\hat{Q}}{2}}}2\sinh{\frac{\hat{P}}{2}} \right)
   \frac{2\sinh{\hat{P}}}{\left( 2\cosh{\frac{\hat{P}}{2}}\right)^{2n}} .
\end{\eq}
By the canonical transformation
\begin{\eq}
(P,Q) \rightarrow (Q,-P) ,
\end{\eq}
we obtain
\begin{\eq}
\hat{\rho}_{\hat{D}_n}
=
 \frac{2\sinh{\hat{Q}}  }{\left( 2\cosh{\frac{\hat{Q}}{2}}\right)^{2n}  } 
           \left( 2\sinh{\frac{\hat{Q}}{2}} \frac{1}{2\cosh{\frac{\hat{P}}{2}}} 2\cosh{\frac{\hat{Q}}{2}}  
                + 2\cosh{\frac{\hat{Q}}{2}} \frac{1}{2\cosh{\frac{\hat{P}}{2}}} 2\sinh{\frac{\hat{Q}}{2}}  
\right) .
\end{\eq}
Indeed this gives 
the same partition function as the $USp+A$ theory with $N_f =n$
because $\hat{\rho}$ of the $USp+A$ theory:
\begin{\eq}
\hat{\rho}_{USp+A} (\hat{Q} ,\hat{P})
=\frac{2\cosh{\hat{Q}}  }  {\left( 4\cosh^2{\frac{\hat{Q}}{2}} \right)^{n -1 } } 
                 \frac{1}{2\cosh{\frac{\hat{P}}{2}}} ,
\end{\eq}
satisfies
\begin{\eqa}
\hat{\rho}_{\hat{D}_n} \frac{1-\hat{R}}{2}
&=&  \frac{2\sinh{\hat{Q}}  }{\left( 2\cosh{\frac{\hat{Q}}{2}}\right)^{2n}  } 
           \left( 2\sinh{\frac{\hat{Q}}{2}} \frac{1-\hat{R}}{2\cosh{\frac{\hat{P}}{2}}} 2\cosh{\frac{\hat{Q}}{2}}  
                + \frac{\cosh^2{\frac{\hat{Q}}{2}}}{\sinh{\frac{\hat{Q}}{2}}} \frac{1-\hat{R}}{2\cosh{\frac{\hat{P}}{2}}} 2\cosh{\frac{\hat{Q}}{2}}  \right) \NN\\
&=&    \frac{1}{2\cosh{\frac{\hat{Q}}{2}}} \left(  \hat{\rho}_{USP+A}  \frac{1-\hat{R}}{2} \right)  2\cosh{\frac{\hat{Q}}{2}} .
\end{\eqa}
Thus, combining the results in app.~\ref{sec:A3D3proof}, app.~\ref{sec:mirrorA}
and app.~\ref{sec:mirrorUSp},
we prove \eqref{UequalUSp}.

\providecommand{\href}[2]{#2}\begingroup\raggedright\endgroup


\begin{thebibliography}{10}

\bibitem{Pestun:2007rz}
V.~Pestun, {\it {Localization of gauge theory on a four-sphere and
  supersymmetric Wilson loops}},  {\em Commun.Math.Phys.} {\bf 313} (2012)
  71--129, [\href{http://xxx.lanl.gov/abs/0712.2824}{{\tt arXiv:0712.2824}}].

\bibitem{Kapustin:2009kz}
A.~Kapustin, B.~Willett, and I.~Yaakov, {\it {Exact Results for Wilson Loops in
  Superconformal Chern-Simons Theories with Matter}},  {\em JHEP} {\bf 1003}
  (2010) 089, [\href{http://xxx.lanl.gov/abs/0909.4559}{{\tt
  arXiv:0909.4559}}];
D.~L. Jafferis, {\it {The Exact Superconformal R-Symmetry Extremizes Z}},  {\em
  JHEP} {\bf 1205} (2012) 159, [\href{http://xxx.lanl.gov/abs/1012.3210}{{\tt
  arXiv:1012.3210}}];
N.~Hama, K.~Hosomichi, and S.~Lee, {\it {Notes on SUSY Gauge Theories on
  Three-Sphere}},  {\em JHEP} {\bf 1103} (2011) 127,
  [\href{http://xxx.lanl.gov/abs/1012.3512}{{\tt arXiv:1012.3512}}].

\bibitem{Marino:2011eh}
M.~Marino and P.~Putrov, {\it {ABJM theory as a Fermi gas}},  {\em
  J.Stat.Mech.} {\bf 1203} (2012) P03001,
  [\href{http://xxx.lanl.gov/abs/1110.4066}{{\tt arXiv:1110.4066}}].

\bibitem{Hatsuda:2013oxa}
Y.~Hatsuda, M.~Marino, S.~Moriyama, and K.~Okuyama, {\it {Non-perturbative
  effects and the refined topological string}},
  \href{http://xxx.lanl.gov/abs/1306.1734}{{\tt arXiv:1306.1734}}.

\bibitem{Matsumoto:2013nya}
S.~Matsumoto and S.~Moriyama, {\it {ABJ Fractional Brane from ABJM Wilson
  Loop}},  {\em JHEP} {\bf 1403} (2014) 079,
  [\href{http://xxx.lanl.gov/abs/1310.8051}{{\tt arXiv:1310.8051}}].

\bibitem{Honda:2014npa}
M.~Honda and K.~Okuyama, {\it {Exact results on ABJ theory and the refined
  topological string}},  {\em JHEP} {\bf 1408} (2014) 148,
  [\href{http://xxx.lanl.gov/abs/1405.3653}{{\tt arXiv:1405.3653}}].

\bibitem{Hatsuda:2013yua}
Y.~Hatsuda, M.~Honda, S.~Moriyama, and K.~Okuyama, {\it {ABJM Wilson Loops in
  Arbitrary Representations}},  {\em JHEP} {\bf 1310} (2013) 168,
  [\href{http://xxx.lanl.gov/abs/1306.4297}{{\tt arXiv:1306.4297}}].

\bibitem{Aharony:2008ug}
O.~Aharony, O.~Bergman, D.~L. Jafferis, and J.~Maldacena, {\it {N=6
  superconformal Chern-Simons-matter theories, M2-branes and their gravity
  duals}},  {\em JHEP} {\bf 0810} (2008) 091,
  [\href{http://xxx.lanl.gov/abs/0806.1218}{{\tt arXiv:0806.1218}}].

\bibitem{Aharony:2008gk}
O.~Aharony, O.~Bergman, and D.~L. Jafferis, {\it {Fractional M2-branes}},  {\em
  JHEP} {\bf 0811} (2008) 043, [\href{http://xxx.lanl.gov/abs/0807.4924}{{\tt
  arXiv:0807.4924}}].

\bibitem{Marino:2009jd}
M.~Marino and P.~Putrov, {\it {Exact Results in ABJM Theory from Topological
  Strings}},  {\em JHEP} {\bf 1006} (2010) 011,
  [\href{http://xxx.lanl.gov/abs/0912.3074}{{\tt arXiv:0912.3074}}];
N.~Drukker, M.~Marino, and P.~Putrov, {\it {From weak to strong coupling in
  ABJM theory}},  {\em Commun.Math.Phys.} {\bf 306} (2011) 511--563,
  [\href{http://xxx.lanl.gov/abs/1007.3837}{{\tt arXiv:1007.3837}}].

\bibitem{Drukker:2009hy}
N.~Drukker and D.~Trancanelli, {\it {A Supermatrix model for N=6 super
  Chern-Simons-matter theory}},  {\em JHEP} {\bf 1002} (2010) 058,
  [\href{http://xxx.lanl.gov/abs/0912.3006}{{\tt arXiv:0912.3006}}];
A.~Klemm, M.~Marino, M.~Schiereck, and M.~Soroush, {\it {ABJM Wilson loops in
  the Fermi gas approach}},  \href{http://xxx.lanl.gov/abs/1207.0611}{{\tt
  arXiv:1207.0611}};
A.~Grassi, J.~Kallen, and M.~Marino, {\it {The topological open string
  wavefunction}},  {\em Commun.Math.Phys.} {\bf 338} (2015), no.~2 533--561,
  [\href{http://xxx.lanl.gov/abs/1304.6097}{{\tt arXiv:1304.6097}}].


\bibitem{Herzog:2010hf}
C.~P. Herzog, I.~R. Klebanov, S.~S. Pufu, and T.~Tesileanu, {\it {Multi-Matrix
  Models and Tri-Sasaki Einstein Spaces}},  {\em Phys.Rev.} {\bf D83} (2011)
  046001, [\href{http://xxx.lanl.gov/abs/1011.5487}{{\tt arXiv:1011.5487}}].

\bibitem{Fuji:2011km}
H.~Fuji, S.~Hirano, and S.~Moriyama, {\it {Summing Up All Genus Free Energy of
  ABJM Matrix Model}},  {\em JHEP} {\bf 1108} (2011) 001,
  [\href{http://xxx.lanl.gov/abs/1106.4631}{{\tt arXiv:1106.4631}}].

\bibitem{Okuyama:2011su}
K.~Okuyama, {\it {A Note on the Partition Function of ABJM theory on $S^3$}},
  {\em Prog.Theor.Phys.} {\bf 127} (2012) 229--242,
  [\href{http://xxx.lanl.gov/abs/1110.3555}{{\tt arXiv:1110.3555}}].

\bibitem{Hanada:2012si}
M.~Hanada, M.~Honda, Y.~Honma, J.~Nishimura, S.~Shiba, and Y.~Yoshida, {\it
  {Numerical studies of the ABJM theory for arbitrary N at arbitrary coupling
  constant}},  {\em JHEP} {\bf 1205} (2012) 121,
  [\href{http://xxx.lanl.gov/abs/1202.5300}{{\tt arXiv:1202.5300}}].


\bibitem{Hatsuda:2012hm}
Y.~Hatsuda, S.~Moriyama, and K.~Okuyama, {\it {Exact Results on the ABJM Fermi
  Gas}},  {\em JHEP} {\bf 1210} (2012) 020,
  [\href{http://xxx.lanl.gov/abs/1207.4283}{{\tt arXiv:1207.4283}}];
P.~Putrov and M.~Yamazaki, {\it {Exact ABJM Partition Function from TBA}},
  {\em Mod.Phys.Lett.} {\bf A27} (2012) 1250200,
  [\href{http://xxx.lanl.gov/abs/1207.5066}{{\tt arXiv:1207.5066}}].

\bibitem{Hatsuda:2012dt}
Y.~Hatsuda, S.~Moriyama, and K.~Okuyama, {\it {Instanton Effects in ABJM Theory
  from Fermi Gas Approach}},  {\em JHEP} {\bf 1301} (2013) 158,
  [\href{http://xxx.lanl.gov/abs/1211.1251}{{\tt arXiv:1211.1251}}];
F.~Calvo and M.~Marino, {\it {Membrane instantons from a semiclassical TBA}},
  {\em JHEP} {\bf 1305} (2013) 006,
  [\href{http://xxx.lanl.gov/abs/1212.5118}{{\tt arXiv:1212.5118}}].

\bibitem{Awata:2012jb}
H.~Awata, S.~Hirano, and M.~Shigemori, {\it {The Partition Function of ABJ
  Theory}},  {\em Prog. Theor. Exp. Phys.} (2013) 053B04,
  [\href{http://xxx.lanl.gov/abs/1212.2966}{{\tt arXiv:1212.2966}}].


\bibitem{Hatsuda:2013gj}
Y.~Hatsuda, S.~Moriyama, and K.~Okuyama, {\it {Instanton Bound States in ABJM
  Theory}},  {\em JHEP} {\bf 1305} (2013) 054,
  [\href{http://xxx.lanl.gov/abs/1301.5184}{{\tt arXiv:1301.5184}}].


\bibitem{Honda:2013pea}
M.~Honda, {\it {Direct derivation of "mirror" ABJ partition function}},  {\em
  JHEP} {\bf 1312} (2013) 046, [\href{http://xxx.lanl.gov/abs/1310.3126}{{\tt
  arXiv:1310.3126}}].

\bibitem{Honda:2014ica}
M.~Honda and S.~Moriyama, {\it {Instanton Effects in Orbifold ABJM Theory}},
  {\em JHEP} {\bf 08} (2014) 091,
  [\href{http://xxx.lanl.gov/abs/1404.0676}{{\tt arXiv:1404.0676}}].

\bibitem{Moriyama:2014gxa}
S.~Moriyama and T.~Nosaka, {\it {Partition Functions of Superconformal
  Chern-Simons Theories from Fermi Gas Approach}},  {\em JHEP} {\bf 11} (2014)
  164, [\href{http://xxx.lanl.gov/abs/1407.4268}{{\tt arXiv:1407.4268}}];
S.~Moriyama and T.~Nosaka, {\it {ABJM membrane instanton from a pole
  cancellation mechanism}},  {\em Phys. Rev.} {\bf D92} (2015), no.~2 026003,
  [\href{http://xxx.lanl.gov/abs/1410.4918}{{\tt arXiv:1410.4918}}].

\bibitem{Moriyama:2014nca}
S.~Moriyama and T.~Nosaka, {\it {Exact Instanton Expansion of Superconformal
  Chern-Simons Theories from Topological Strings}},  {\em JHEP} {\bf 05} (2015)
  022, [\href{http://xxx.lanl.gov/abs/1412.6243}{{\tt arXiv:1412.6243}}].

\bibitem{Hatsuda:2015lpa}
Y.~Hatsuda, M.~Honda, and K.~Okuyama, {\it {Large N non-perturbative effects in
  $\mathcal{N}=4$ superconformal Chern-Simons theories}},  {\em JHEP} {\bf 09}
  (2015) 046, [\href{http://arxiv.org/abs/1505.07120}{{\tt arXiv:1505.07120}}].

\bibitem{Marino:2012az}
M.~Marino and P.~Putrov, {\it {Interacting fermions and N=2 Chern-Simons-matter
  theories}},  \href{http://xxx.lanl.gov/abs/1206.6346}{{\tt arXiv:1206.6346}}.

\bibitem{Mezei:2013gqa}
M.~Mezei and S.~S. Pufu, {\it {Three-sphere free energy for classical gauge
  groups}},  {\em JHEP} {\bf 02} (2014) 037,
  [\href{http://xxx.lanl.gov/abs/1312.0920}{{\tt arXiv:1312.0920}}].

\bibitem{Assel:2015hsa}
B.~Assel, N.~Drukker, and J.~Felix, {\it {Partition functions of 3d $\hat
  D$-quivers and their mirror duals from 1d free fermions}},  {\em JHEP} {\bf
  08} (2015) 071, [\href{http://arxiv.org/abs/1504.07636}{{\tt
  arXiv:1504.07636}}].

\bibitem{Moriyama:2015jsa}
S.~Moriyama and T.~Nosaka, {\it {Superconformal Chern-Simons Partition
  Functions of Affine D-type Quiver from Fermi Gas}},  {\em JHEP} {\bf 09}
  (2015) 054, [\href{http://arxiv.org/abs/1504.07710}{{\tt arXiv:1504.07710}}].

\bibitem{Bhattacharyya:2012ye}
S.~Bhattacharyya, A.~Grassi, M.~Marino, and A.~Sen, {\it {A One-Loop Test of
  Quantum Supergravity}},  \href{http://xxx.lanl.gov/abs/1210.6057}{{\tt
  arXiv:1210.6057}}.

\bibitem{Dabholkar:2014wpa}
A.~Dabholkar, N.~Drukker, and J.~Gomes, {\it {Localization in supergravity and
  quantum $AdS_4/CFT_3$ holography}},  {\em JHEP} {\bf 10} (2014) 90,
  [\href{http://xxx.lanl.gov/abs/1406.0505}{{\tt arXiv:1406.0505}}].

\bibitem{Imamura:2008nn}
Y.~Imamura and K.~Kimura, {\it {On the moduli space of elliptic
  Maxwell-Chern-Simons theories}},  {\em Prog. Theor. Phys.} {\bf 120} (2008)
  509--523, [\href{http://xxx.lanl.gov/abs/0806.3727}{{\tt arXiv:0806.3727}}].

\bibitem{Gaiotto:2008sd}
D.~Gaiotto and E.~Witten, {\it {Janus Configurations, Chern-Simons Couplings,
  And The theta-Angle in N=4 Super Yang-Mills Theory}},  {\em JHEP} {\bf 06}
  (2010) 097, [\href{http://xxx.lanl.gov/abs/0804.2907}{{\tt
  arXiv:0804.2907}}];
K.~Hosomichi, K.-M. Lee, S.~Lee, S.~Lee, and J.~Park, {\it {N=4 Superconformal
  Chern-Simons Theories with Hyper and Twisted Hyper Multiplets}},  {\em JHEP}
  {\bf 07} (2008) 091, [\href{http://xxx.lanl.gov/abs/0805.3662}{{\tt
  arXiv:0805.3662}}].

\bibitem{Grassi:2014vwa}
A.~Grassi and M.~Marino, {\it {M-theoretic matrix models}},  {\em JHEP} {\bf
  02} (2015) 115, [\href{http://xxx.lanl.gov/abs/1403.4276}{{\tt
  arXiv:1403.4276}}].

\bibitem{Hatsuda:2014vsa}
Y.~Hatsuda and K.~Okuyama, {\it {Probing non-perturbative effects in
  M-theory}},  {\em JHEP} {\bf 10} (2014) 158,
  [\href{http://xxx.lanl.gov/abs/1407.3786}{{\tt arXiv:1407.3786}}].

\bibitem{Okuyama:2015auc}
K.~Okuyama, {\it {Probing non-perturbative effects in M-theory on
  orientifolds}},  \href{http://arxiv.org/abs/1511.02635}{{\tt
  arXiv:1511.02635}}.

\bibitem{Gulotta:2012yd}
D.~R. Gulotta, J.~Ang, and C.~P. Herzog, {\it {Matrix Models for Supersymmetric
  Chern-Simons Theories with an ADE Classification}},  {\em JHEP} {\bf 1201}
  (2012) 132, [\href{http://xxx.lanl.gov/abs/1111.1744}{{\tt
  arXiv:1111.1744}}];
D.~R. Gulotta, C.~P. Herzog, and T.~Nishioka, {\it {The ABCDEF's of Matrix
  Models for Supersymmetric Chern-Simons Theories}},  {\em JHEP} {\bf 1204}
  (2012) 138, [\href{http://xxx.lanl.gov/abs/1201.6360}{{\tt
  arXiv:1201.6360}}];
  P.~M. Crichigno, C.~P. Herzog, and D.~Jain, {\it {Free Energy of $D_n$ Quiver
    Chern-Simons Theories}},  {\em JHEP} {\bf 03} (2013) 039,
    [\href{http://xxx.lanl.gov/abs/1211.1388}{{\tt arXiv:1211.1388}}].

\bibitem{Moriyama:2015asx}
S.~Moriyama and T.~Suyama, {\it {Instanton Effects in Orientifold ABJM
  Theory}},  \href{http://arxiv.org/abs/1511.01660}{{\tt arXiv:1511.01660}}.

\bibitem{Hosomichi:2008jb}
K.~Hosomichi, K.-M. Lee, S.~Lee, S.~Lee, and J.~Park, {\it {N=5,6
  Superconformal Chern-Simons Theories and M2-branes on Orbifolds}},  {\em
  JHEP} {\bf 09} (2008) 002, [\href{http://xxx.lanl.gov/abs/0806.4977}{{\tt
  arXiv:0806.4977}}].

\bibitem{Grassi:2014uua}
A.~Grassi, Y.~Hatsuda, and M.~Marino, {\it {Quantization conditions and
  functional equations in ABJ(M) theories}},
  \href{http://xxx.lanl.gov/abs/1410.7658}{{\tt arXiv:1410.7658}}.

\bibitem{Codesido:2014oua}
S.~Codesido, A.~Grassi, and M.~Marino, {\it {Exact results in $ \mathcal{N}=8 $
  Chern-Simons-matter theories and quantum geometry}},  {\em JHEP} {\bf 07}
  (2015) 011, [\href{http://xxx.lanl.gov/abs/1409.1799}{{\tt
  arXiv:1409.1799}}].

\bibitem{Cheon:2012be}
S.~Cheon, D.~Gang, C.~Hwang, S.~Nagaoka, and J.~Park, {\it {Duality between N=5
  and N=6 Chern-Simons matter theory}},  {\em JHEP} {\bf 11} (2012) 009,
  [\href{http://xxx.lanl.gov/abs/1208.6085}{{\tt arXiv:1208.6085}}].

\bibitem{Intriligator:1996ex}
K.~A. Intriligator and N.~Seiberg, {\it {Mirror symmetry in three-dimensional
  gauge theories}},  {\em Phys.Lett.} {\bf B387} (1996) 513--519,
  [\href{http://xxx.lanl.gov/abs/hep-th/9607207}{{\tt hep-th/9607207}}];
A.~Hanany and E.~Witten, {\it {Type IIB superstrings, BPS monopoles, and
  three-dimensional gauge dynamics}},  {\em Nucl.Phys.} {\bf B492} (1997)
  152--190, [\href{http://xxx.lanl.gov/abs/hep-th/9611230}{{\tt
  hep-th/9611230}}].

\bibitem{deBoer:1996mp}
J.~de~Boer, K.~Hori, H.~Ooguri, and Y.~Oz, {\it {Mirror symmetry in
  three-dimensional gauge theories, quivers and D-branes}},  {\em Nucl. Phys.}
  {\bf B493} (1997) 101--147,
  [\href{http://xxx.lanl.gov/abs/hep-th/9611063}{{\tt hep-th/9611063}}].


\bibitem{Assel:2014awa}
B.~Assel, {\it {Hanany-Witten effect and SL(2, $\mathbb{Z}$) dualities in
  matrix models}},  {\em JHEP} {\bf 10} (2014) 117,
  [\href{http://xxx.lanl.gov/abs/1406.5194}{{\tt arXiv:1406.5194}}].

\bibitem{Drukker:2015awa}
N.~Drukker and J.~Felix, {\it {3d mirror symmetry as a canonical
  transformation}},  {\em JHEP} {\bf 05} (2015) 004,
  [\href{http://arxiv.org/abs/1501.02268}{{\tt arXiv:1501.02268}}].

\bibitem{Moriyama:2016xin}
S.~Moriyama and T.~Suyama, {\it {Orthosymplectic Chern-Simons Matrix Model and
  Chirality Projection}},  \href{http://arxiv.org/abs/1601.03846}{{\tt
  arXiv:1601.03846}}.

\bibitem{Kapustin:2010xq}
A.~Kapustin, B.~Willett, and I.~Yaakov, {\it {Nonperturbative Tests of
  Three-Dimensional Dualities}},  {\em JHEP} {\bf 1010} (2010) 013,
  [\href{http://xxx.lanl.gov/abs/1003.5694}{{\tt arXiv:1003.5694}}].

\end{thebibliography}

\end{document}